\documentclass[a4paper, 12pt]{article}
\usepackage[margin=2.5cm]{geometry}
\usepackage{setspace,parskip}
\usepackage{colortbl}
\usepackage[dvipsnames]{xcolor}
\newcommand\crule[3][black]{\textcolor{#1}{\rule{#2}{#3}}}
\usepackage{natbib}
\usepackage{url,hyperref}

\usepackage{xcolor}
\definecolor{EconomicsGray}{RGB}{198,212,225}
\definecolor{EconomicsLightBlue}{RGB}{127,191,192}
\definecolor{EconomicsBlue}{RGB}{0,63,138}
\definecolor{EconomicsDarkBlue}{RGB}{0,63,117}
\hypersetup{
	colorlinks,
	linkcolor=EconomicsDarkBlue,
	citecolor=EconomicsBlue,
	urlcolor={blue!50!black}
}
\newcommand*{\doi}[1]{\href{http://dx.doi.org/#1}{\nolinkurl{#1}}}
\usepackage{graphicx}
\usepackage{amsmath}
\usepackage{amsthm,booktabs, array} 
\newcolumntype{L}[1]{>{\raggedright\arraybackslash}b{#1}}
\usepackage{subcaption}
\DeclareCaptionLabelSeparator{pointiret}{ \textbar$  $ }
\usepackage[labelfont={bf }, labelsep=period]{caption}
\usepackage{algorithm2e} \SetAlgoCaptionSeparator{.}

\usepackage[british]{babel}
\usepackage{mathptmx} 
\usepackage[printonlyused]{acronym} 
\usepackage{indentfirst} 
\usepackage{multirow}
\usepackage{bm}
\usepackage{titlesec}

\titleformat{\section}    
{\normalfont\fontfamily{phv}\fontsize{14}{14}\bfseries}{\thesection}{1.5em}{}
\titleformat{\subsection}    
{\normalfont\fontfamily{phv}\fontsize{14}{17}\bfseries}{\thesubsection}{1.5em}{}
\title{Bayesian Sample Size Determination for Planning Hierarchical Bayes Small Area Estimates}

\date{\footnotesize \sffamily Manuscript draft v1.1 \\ 18 January 2018}

\author{Peter Dutey-Magni$^{1,2} $ \\
	{\footnotesize $ ^1 $UCL Institute of Health Informatics,  University College London, 222 Euston Road, London, NW1 2DA}\\
	{\footnotesize $ ^2 $Faculty of Medicine, University of Southampton, 4-12 Terminus Terrace, Southampton, SO14 3DT}\\
	{\footnotesize \texttt{p.dutey-magni@ucl.ac.uk}}}

\begin{document}
	\maketitle 

\begin{abstract}
This paper devises a fully Bayesian sample size determination method for hierarchical model-based small area estimation with a decision risk approach. A new loss function specified around a desired maximum posterior variance target implements conventional official statistics criteria of estimator reliability (coefficient of variation of up to 20 per cent). This approach comes with an efficient binary search algorithm identifying the minimum effective sample size needed to produce small area estimates under this threshold constraint. Traditional survey sampling design tools can then be used to plan appropriate data collection using the resulting effective sample size target. This approach is illustrated in a case study on small area prevalence of life limiting health problems for 6 age groups across 1,956 small areas in Northern England, using the recently developed Integrated Nested Laplace Approximation method for spatial generalised linear mixed hierarchical models. \bigskip

\end{abstract}

\section{Introduction}
When small area estimates from survey samples are imprecise, model-based estimation is often used to borrow strength and produce more efficient estimators. Hierarchical Bayes (\acsu{HB}) prediction, in particular, has been shown to yield good results and make efficient use of available data \citep{Ghosh1994b}. When it comes to routine applications, an question commonly faced by practitioners is how to determine the minimum sample size necessary to achieve a desired precision, generally expressed as a target standard error or \ac{RSE} applicable to all target small areas. Although some analytical solutions have been proposed in the model-assisted literature, this problem has remained relatively unexamined in model-dependent \ac{SAE}.

This paper provides a fully Bayesian treatment of this problem. Section \ref{p3:sec:review} briefly reviews existing work on Bayesian \ac{SSD} for clinical trials and other investigations, and draw implications for planning model-based \ac{SAE} studies. Section \ref{p3:sec:methods} describes a model-based simulation procedure to determine \ac{ESS} requirements under a maximum relative posterior variance constraint. This approach is applied in section \ref{p3:sec:casestudy} using 2011 UK census data on chronic health and a spatial hierarchical generalised linear mixed model. A design-based simulation study confirms both (a) the validity of solution produced by the \ac{SSD} algorithm and (b) efficiency savings achieved over traditional survey sampling direct estimation in section \ref{p3:sec:results}. Finally, section \ref{p3:sec:conclusion} discusses generalisation to other model-based methods as well as more complex sampling designs.

\section{Review of \acs{SSD} for hierarchical models}\label{p3:sec:review}
In the frequentist approach, target sample sizes are usually determined by reference to the sampling distribution of the target parameter under a given survey sampling plan. With complex model-assisted or model-dependent statistical designs, this sampling distribution is typically unknown. This is particularly the case when a study examines a model parameter estimated under a multivariate statistical model with covariate inputs.

Amongst those complex statistical designs are two-stage hierarchical models commonly used to estimate some kind of effect in multicentre randomised controlled trials or in multilevel observational studies (e.g. studies of educational attainment across classes and schools). This effect can be a treatment effect, a regression coefficient or some other cluster-specific characteristic. We distinguish between two categories of studies based on their inferential motivation:
\begin{enumerate}
	\item studies aiming to detect a \textit{global effect} with a predefined statistical power;
	\item studies aiming to estimate \textit{effect sizes} within each centre/cluster with a predefined precision.
\end{enumerate}

\subsection{\acs{SSD} for detection}\label{p3:sec:review:cat1}
In category (1), success is defined as collecting a sufficient number of observations across a sufficient number of centres or clusters so that the target statistical power is achieved in the model of interest. Achieving the target statistical power allows study investigators to reject a null hypothesis if the size of the treatment effect exceeds a clinically meaningful threshold set at the design stage. In observational study designs, examples are varied which optimise, for instance, the number of clusters and the number of units within clusters. Several software applications now exist across the educational, behavioural and wider multilevel literature \citep{Snijders2005, Cools2008,Browne2009, Zhang2009a}. In interventional research, including clinical trials, some design problems have closed-form solutions \citep{Raudenbush1997}, while others are more complex and require computer-intensive Monte Carlo experiments to reach a solution. \citet{Joseph1995} proposed algorithms relying on a binary sample size search between 0 and the frequentist binomial sample size requirement to optimise the \ac{SSD} solution at a reasonable computational cost. The adequate sample size is determined as the smallest sample size meeting the chosen constraints.

\subsection{\acs{SSD} for prediction}\label{p3:sec:review:cat2}
The present paper is concerned with studies belonging to category (2), in which success is defined as collecting a sufficient number of observations to estimate the treatment effect with a predefined precision within each centre or cluster. In contrast with category (1), research on \ac{SSD} for a vector parameter estimated under a working model is not widely developed, with the exception of clinical audit design methodology. In a series of simulation studies, \citet{Normand2002} explored the effect of the number of clusters and cluster sample sizes on the efficiency of audits under beta-binomial hierarchical models borrowing strength and adjusting for confounders. \citet{Zou2001} and \citet{Zou2003} proposed an \ac{SSD} algorithm in three-stage hierarchical models search for an design for hospital benchmarking clinical audits. In these examples, the chosen \ac{SSD} criterion is an upper bound for the average width of posterior intervals of centre-level estimates. Their method involves a grid search of the minimum target sample size (a sample size sweep in a predetermined sequence of candidate values). It is implemented with Monte Carlo simulation using the prespecified model, successively: sampling from the model  priors; simulating a realisation from the model; predicting the corresponding parameter; and computing the level of compliance with the target posterior interval width. This type of Monte Carlo solution associated with Bayesian decision theory are attracting growing interest for complex \ac{SSD} problems in hierarchical designs for medical studies. Another area for application is the design of spatial sampling. Even with the simplest working models, \citet{Diggle2006} conclude that sampling simulations are inevitable, albeit computationally intensive. The authors nevertheless recognise that the latest model estimation methods are likely to make the simulation approach feasible.

\subsection{\acs{SSD} for small area studies}
\ac{SSD} techniques have scarcely been applied to plan model-based \ac{SAE} studies, that is when a single working model is used to compute predictions within each small area in a given study population. Although simulation studies are ubiquitous in the \ac{SAE} literature, they have been used either to illustrate efficiency gains obtained from more complex modelling designs \citep{Jonker2013a, Porter2015, Ross2015}, or to validate a small area model against historical data \citep{Barker2013}.
Closed-form \ac{SSD} solutions are only available for the simplest models: see \citet{Falorsi2008, Molefe2011,Molefe2015} in the model-assisted literature, and \citet{Raudenbush1997} who looks at implications of introducing a model covariate on \ac{SSD}. Just recently, \citet{Keto2017} proposed solutions for optimal sample allocation across small areas in relation to the Empirical Best Linear Unbiased Prediction. Yet, the availability of \ac{SSD} methods for real-world statistical needs remains of particular importance with model-based estimation since precision depends not just on the type of predictor selected, but on covariates and random components included in the model \citep{Rao2011a, Rotondi2009}. The \ac{MSE} of predictors depends not solely on a known sampling distribution, but also on the working model's covariance structure. To the best of our knowledge, not \ac{SSD} methodology is available for emerging developments such as spatial smoothing models \citep{Pratesi2008d, Gomez-Rubio2010}, or small area models borrowing strength from time \citep{You2011}, age or cohort effects \citep{Congdon2006c}.

\subsection{Prerequisites of frequentist and Bayesian solutions for \acs{SSD}}
\ac{SSD} presupposes (1) design constraints in the form of a set of criteria against which to optimise sample design; and (2) prior information on the population of study.

The implementation of design constraints is straightforward in closed-form frequentist solutions which focus on estimator variance targets. In fully Bayesian \ac{SSD}, design criteria are often treated as a decision rule determined by a loss function \citep{Adcock1988,Adcock1997}. A typical decision rule for studies preoccupied with hypothesis testing is based on a function of statistical power (opposite of the false negative rate) and significance level (opposite of the false positive error rate), as illustrated by \citet{Sahu2006}. Yet other rules have been proposed for detection studies; \citet{Joseph1995} considered three such design objectives: achieving a desired average coverage probability for highest posterior density intervals; satisfying a maximum average length for those intervals; and a combination of the two constraints. As for studies interested in prediction, decision rules have in majority been specified in relation to the width of interval estimates; see for instance \citet{Joseph1995} and \citet{Zou2001}. These overlap conventional government survey design criteria defined in terms of precision, either based on an estimator's margin of error or \ac{RSE}. For instance the 2000/01 English Local boost of the Labour Force Survey was designed with a frequentist approach to insure an acceptable precision, defined as a maximum \ac{RSE} of 20 per cent for design-based estimators of economic activity headcounts of 6,000 districts \citep[p.~40]{Hastings2002}. A model-based equivalent is the relative root \ac{MSE}, while a fully Bayesian equivalent would be to consider a function of the relative posterior variance, for instance the number of  districts which fail to meet a maximum relative posterior variance threshold.

The second prerequisite of \ac{SSD}, namely prior information around the parameter of interest, is treated very differently in the frequentist and the Bayesian apparach \citep{Adcock1997}. In the frequentist paradigm, \ac{SSD} is entirely determined by the sampling distribution of the study target parameter, which itself depends on its population variance, which is typically unknown. In the absence of knowledge regarding the population variance of the study parameter, the standard frequentist approach to \ac{SSD} involves plugging an assumed value for this variance in a closed formula. The outcome is therefore entirely dependent on how conservative this assumption is and it is often necessary to overestimate the population variance.
In contrast, fully Bayesian \ac{SSD} does not handle unknown parameters using a plug-in method but instead using explicit priors and hyperpriors \citep{Adcock1997}. The literature contains a variety of examples. When determining a sample size for a binomial proportion, \citet{Joseph1995} and \citet{Zou2001} set the scale and rate (hyperparameters) of the beta distribution (prior) believed to determine an overdispersed binomial distribution of interest. In biomedical research, such hyperpriors are generally elicited from pilot data, previous studies and subjective expert opinion \citep{Spiegelhalter1986}.

At the first glance, particularly when relying on pilot or historical data to form a prior, it seems intuitive to use a single prior for both the design and the analysis. In other words, the set of priors used to simulate prediction under various sample size scenarios is also incorporated in the working model. Yet  \citet{Spiegelhalter1986} and \citet{Sahu2006}  
have argued in favour of separating design and fitting priors on scientific grounds. Regulations on biomedical trials sometimes impose that the data are analysed under a state of pre-experimental knowledge, that is without incorporating knowledge from data produced in previous studies. Although such historical data can be valuable in optimising \ac{SSD}, it is not necessarily desirable to introduce them in the analysis itself as this can be left to subsequent meta-analytic studies. Similar constraints may exist in official statistics, where the reliance on informative priors is sometimes subject to objections \citep{Fienberg2011}. When the only available source of prior knowledge is historical data, there may be reasons to restrict its use to \ac{SSD}. This provides further assurance in the solution of \ac{SSD} while determining sample requirements to produce a sufficiently precise estimate without having to pool data from previous statistical bulletins. This is by no mean the only way to proceed, but it is sometimes desirable to treat the elicitation of design priors and fitting priors separately. 

On the one hand, design priors retain a strong influence on the outcome of any \ac{SSD} procedure and its success. Particular attention has been given to robust prior elicitation---that is priors that do not convey excessive confidence compared to the existing knowledge, and which are flexible enough to offer protection against misspecified models. 
With regard to \ac{SAE}, priors can be elicited from previous survey waves or pilots: routine government survey data are typically abundant. In principle, the most simple type of design prior can consist of hyperparameters of random components since they determine the level of shrinkage in \ac{HB} prediction and, by way of consequence, posterior variance. Such hyperpriors can be elicited from appropriate marginal posteriors obtained from the combination of pilot data with a vague uninformative prior (which can be the fitting prior). Due consideration must be given to how much belief can be placed into the stability of these parameters across years or across surveys, and it may be necessary to apply a small discount to their influence (see \citealt{DeSantis2007}).

On the other hand, fitting priors can be everything between vague and informative. The use of uninformative hyperpriors for random components is not always an option as it often leads to difficulties in estimating models. Both Markov chains Monte Carlo and \ac{INLA} encounter numerical difficulties with complex models, especially with spatial models.  \citet{Fong2010} suggest specifying weakly informative priors for Gaussian random effects by using the log Student \textit{t} distribution (with one or two degrees of freedom) and predefined lower and upper bounds for the range of 95 per cent of realisations. Hyperpriors can then be deduced in the form of the scale and rate of a Gamma distribution. More recently, \citet{Simpson2014} have addressed more complex models with a combination of random effects and proposed a weakly informative `penalised complexity prior' based on some belief of the scale of random effect (standard deviation).

\section{Model-based \acs{SSD}} \label{p3:sec:methods}
We consider the estimation of a small area characteristic (e.g. economic status, illness, income, marital status) under a given working model $ \mathcal{M} $.
Let the population $ U $ be partitioned into small areas $ d=\{1,\ldots, D\}$. $\mathbf{N} = \{N_{1},\ldots, N_{D}\}$ and  $\mathbf{Y} = \{ Y_{1},\ldots, Y_{D}\}$ respectively denote the population size in area $ d $ and the characteristic total in area $ d $: this can be the area headcount of  individuals with the given characteristic, or the area total (such as total income). We are interested in estimating population means $ \overline{Y}_{d}= Y_{d}/N_{d}$ using
\begin{itemize}
	\item hierarchical working model $ \mathcal{M} $, to produce an \ac{HB} predictor of  $ \overline{Y}_{d}$ notated $ \theta_{d} $;
	\item data from auxiliary covariates $ \mathbf{X} $ available for the entire population;
	\item a sample survey $ s $ to be designed.
\end{itemize}

We seek to determine $ f $, the effective sampling fraction for $ s $, using  Bayesian \ac{SSD} under some design constraints. We remark that area-specific \acp{ESS} $ n_{d}$ are such that $ n_{d} \sim \text{Binomial}(N_{d}, f) $.

\subsection{Sample size criteria and Bayes decision rule}\label{p3:sec:methods:criteria}
\def\krul{\leavevmode\hbox{\lower2.5pt\vbox to10pt{}\kern1.8pt
		\pdfliteral{q   
			1 j .7 0 0 .7 0 0 cm
			-2 1 m 
			3 2 4 5 4 7  c
			4 9 3 10 2 10 c
			1 10 0 9 0 7  c
			0 3 5 2 5 0 c
			5 -2 2 -3 1 -3 c
			S Q}\kern4.5pt}}

A conventional frequentist criterion of statistical reliability in official statistics is the \ac{RSE} or \ac{CV}. A possible Bayesian equivalent is the relative posterior variance of the \ac{HB} predictor $ \theta_d $. Tabular cells of estimates with a relative posterior variance in excess of 20 per cent are to be suppressed from statistical publications. 
It is desirable that the overall rate of cell suppression remains low, for example below a threshold $ \krul = 0.01 $. We implement this requirement through a simple design loss function: the proportion of cell suppression in the dataset weighted by the population headcount of those cells. The total loss $\ell(f \ | \ \ldots) $ can be thought of as the total headcount of populations eligible to a reliable estimate, whose estimates have to suppressed due to reliability concerns. The weighting introduces a form of trade-off, which prioritises reliable estimates for large populations while being more tolerant of the risk of cell suppression for the smallest cross-classifications, which are inevitably the most demanding in terms of data collection.
\begin{equation}\label{p3:lossfunction}
\ell(f \ | \ \mathcal{M},  \pi(\tau_\gamma), \pi(\tau_\upsilon), \pi(\tau_\nu), \ldots) =   N^{-1} \sum_{d}\left[   N_{d} \ I(\text{RSE}(\theta_{d}) > 0.2)\right] 
\end{equation}

In this expression, $ I(\cdot) $ is the indicator function. The overall loss is an intractable function of the sampling fraction $ f $ and is conditional on both the working model $ \mathcal{M} $ and the set of design and fitting priors $ \pi(\cdot) $. Though $\ell$ has no obvious closed-form expression, it is reasonable to take the premise that it is a monotonically decreasing function of $ n $ (for a discussion of the design consistency of \ac{HB} prediction, see \citealt{Lahiri2007c}). This facilitates the evaluation of integral $ \int \ell(f \ | \ \ldots) \ \ df $ over a reasonable interval of sampling fractions $ [a,b] $.

Many more specifications can be considered for the loss. $ \ell $ can be defined with respect to the width of prediction intervals such as highest posterior density intervals rather than \ac{RSE}. It can also be restricted to a subset of domains $ d $ in the event that quality standards set by statutory or funding requirements do not apply to all domains $ d $. 

Depending on design constraints and costs, it is possible to envisage more sophisticated loss functions inspired from optimal sample design or the `value of information' approach. In particular, it is conceivable to attribute a price to cell suppressions or to penalised $ \ell $ by a marginal cost function of increasing the sampling fraction. These constitute options for more holistic decision rule and lead to an optimum between cost saving and exhaustive publication. Provided that such refinements strictly depend on $ f $ and posterior means or variances, they add no further dimension to the \ac{SSD} equation, and the approach described in this paper should remain entirely applicable.

\begin{figure}[h]
	\includegraphics[width=.7\textwidth]{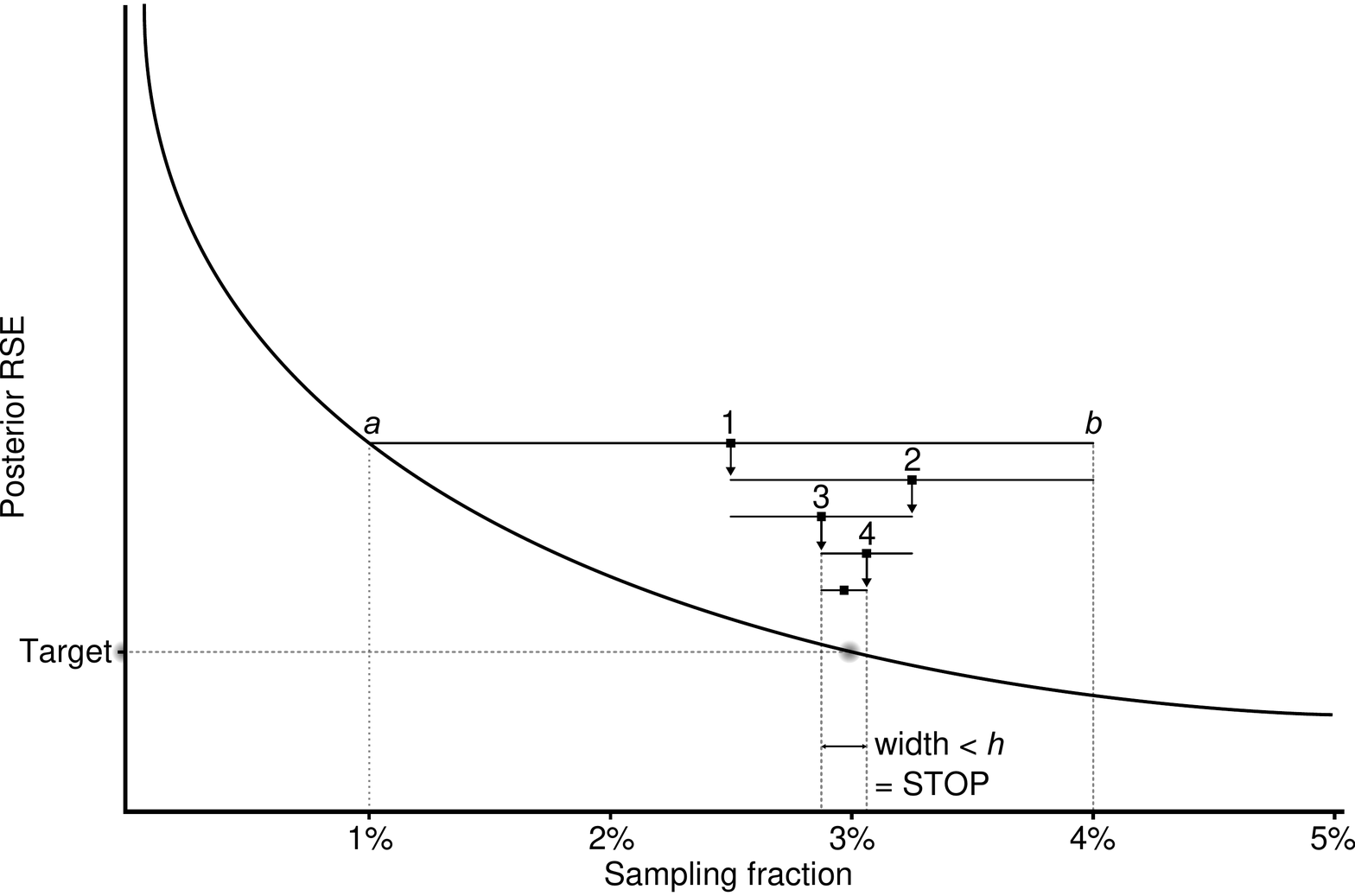}
	\centering
	\caption{Graphical illustration of the binary algorithm} \label{p3:fig:algo1_illustration}
\end{figure}

\begin{figure}
\begin{minipage}{.6\linewidth}  
	\setlength{\algomargin}{1.5em} \RestyleAlgo{boxed} 
	\begin{algorithm}[H] 
		\nl\KwIn{$ s_0 $, $ \pi(\theta_{d} \ | \ s_0) $, $ f_a$, $ f_b$, $ \ell(f_a) $, $ \ell(f_b) $, $ h$, $ \krul $, $ L$, $ \gamma $}
		\nl$k\leftarrow0; l\leftarrow0$\;
		\Repeat{$f_b - f_a < h$}{
			\nl$ k\leftarrow k+1 $\;
			\nl$f_k\leftarrow (f_a+f_b)/2$\;
			\Repeat{l = L}{
				\nl$ l\leftarrow l+1$\;
				\nl sample $ \theta_{d}^\star $ from prior $ \pi(\theta_{d} \ | \ s_0) $\;
				\nl simulate $ n_{dl}^\star \sim \text{Binomial}(N_{d},f_k)$\;
				\nl simulate $  n_{dl}^\star  $ realisations $ \bar{y}_{dl}^\star $ from the likelihood of   $  \theta_{dl}^\star  $\;
				\nl fit model $\mathcal{M}^\star_l$ on data $ \bar{y}_{dl}^\star $ \;
				\nl estimate posterior density $ \pi(\theta_{dl}^\star |y_{dl}^\star, \mathcal{M}^\star_l) $ \;
				\nl compute relative posterior variances $ \widehat{\text{RSE}}(\theta_{dl}^\star) $ \label{p3:algo1:postRSE}\;
			}
			\nl$ \ell(f_k)_l \leftarrow N^{-1}   \sum_{d=1}^{D} \left(   N_{d} \ I\left[ \widehat{\text{RSE}}(\theta_{dl}^\star) > 0.2\right] \right)  $ \label{p3:algo1:lossfunction}\; 
			\uIf{{\normalfont$ \Pr(\ell(f_k)\leq \krul ) < \gamma $}}{
				\nl$ f_b\leftarrow f_k $} 
			\Else{
				\nl$ f_a\leftarrow f_k $}} 
		\nl\KwRet{$f_a, f_b, \ell(a), \ell(b)$}
	\end{algorithm}

\end{minipage}
\centering
\caption{Binary \acs{SSD} algorithm}\label{p3:algo1}
\end{figure}
\subsection{Sample size minimisation}\label{p3:sec:methods:essminimisation}
Simulating the relative posterior variance for all possible sample sizes until the minimum quality standard is attained following a sweep approach is computationally cumbersome. We instead opt for a binary search algorithm (see Figures \ref{p3:fig:algo1_illustration} and \ref{p3:algo1}) to minimise $ \ell(f) $. By progressing iteratively towards the solution with steps of decreasing sizes, we considerably reduce the number of simulations. The algorithm starts with $ f_1 $, the midpoint of the interval $ [a,b] $, and evaluates $ \ell(f_1) $ over many replications of the sampling and model estimation process. This process is then replicated on interval $[a, f_1] $ if $\ell(f_1)$ can be trusted to fall below the maximum tolerated suppression setting $ \krul $ or $ [f_1,b] $ in the opposite case---that is, if more than $ \krul $ per cent of sampling simulations return a strictly positive value for $ \ell(f_1) $. The interval of possible solutions $ f_k $ is therefore halved repeatedly until the interval width is small enough to be used; this is determined by a tolerance setting $ h $. The upper bound of the final interval is the target \ac{ESS} solution. The algorithm can be restarted on the final interval to reach an even narrower $ h $. By construct, the number of iterations required to complete the algorithm  is the smallest integer  $ k_{\text{max}} $ such that $ k_{\text{max}} > (2h)^{-1}(b-a)  $.

It is worth emphasising that this procedure relies on $ \ell $ being a monotonous (not strictly) decreasing function of sampling fraction $f$. The algorithm progression is mainly determined by a risk setting $ \gamma $, the probability that cell suppression will exceed $ \krul $ for any given $ f_k $. The number of simulations $ L $ ordered for each sampling fraction $ f_k $ should be set in relation to $ \gamma $; we suggest that the order of magnitude of $ L $ should always be at least equal to the order of magnitude of $\gamma^{-1} $.

\section{Case study} \label{p3:sec:casestudy}
\subsection{Context}
The approach described in section \ref{p3:sec:methods} is illustrated on a typical tabular Office for National Statistics census output. This case study examines sample size requirements and efficiency gains under small area models predicting the age-specific prevalence of  \ac{LLTI} across small areas in Northern England. The UK government having announced its goal to discontinue full population enumeration through traditional decennial censuses \citep{UKCabinet2014}, it is expected that small area population health characteristics will in the future have to be estimated using sample surveys and administrative sources. The aim of the present case study is twofold: (i) to apply the \ac{SSD} procedure set out in section \ref{p3:sec:methods:essminimisation} to reproduce 2011 census health outputs under conventional UK standards of statistical reliability; (ii) to assess efficiency gained from incorporating more advanced hierarchical prior structures into small area models (borrowing strength from neighbouring areas and age groups).

Future census quality standards published by the \citet[p.~44]{Beyond2011.O4} demand a maximum \ac{RSE} of 20 per cent for population headcounts representing at least 3 per cent of an \ac{MSOA}'s population. Assuming an average resident population of 8,000 across \acp{MSOA}, this criterion is equivalent to a margin of error of up to $ \pm 94 $ for a headcount of 240 persons. \ac{MSOA} zones were originally designed to ensure lower and upper population limits of 5,700 to 11,100 usual residents. By 2014, 10 per cent of these zones no longer met such limits \citep{ONS2015a}. As a consequence, the future Office for National Statistics' census quality standards in effect require a much smaller margin of error for the smallest \acp{MSOA}.

\subsection{Design constraints on a tabular output}
The case study population $ U $ is partitioned into \acp{MSOA} $ d=\{1,\ldots, D\}$ and age groups $ j =\{1,\ldots, J\}  $. $\mathbf{N} = \{N_{11},\ldots, N_{JD}\}$ and  $\mathbf{Y} = \{ Y_{11},\ldots, Y_{JD}\}$ respectively denote headcounts of members of private households of age group $ i $ residing in area $ d $ and the corresponding number of individuals reporting an \ac{LLTI}. We are interested in the \ac{LLTI} prevalence proportions $ p_{jd} = Y_{jd}/N_{jd}$. Table \ref{p3:tab:tabularexample} illustrates the typical tabular census output corresponding to this data.

\begin{table} \footnotesize \centering
	\caption{Example of tabular census output for health characteristics} \label{p3:tab:tabularexample}
	\begin{tabular}{cccc|ccc|ccccc}
		\toprule
		& \multicolumn{9}{c}{LLTI status}& \\
		\cmidrule(){2-10}
		& \multicolumn{3}{c}{LLTI}& \multicolumn{3}{c}{No LLTI} &\multicolumn{3}{c}{All}&  \\
		\cmidrule(r){2-4} \cmidrule(lr){5-7} \cmidrule(l){8-10}
		&1 &$ \cdots $ & J & 1 &$ \cdots $ & J & 1 &$ \cdots $ & J & Total \\
		\cmidrule(r){2-4} \cmidrule(lr){5-7} \cmidrule(l){8-10} \cmidrule(l){11-11}
		1 & \cellcolor{Goldenrod}$ Y_{11} $ & \cellcolor{Goldenrod}$ \cdots $ &\cellcolor{Goldenrod} $ Y_{J1} $&$ N_{11}-Y_{11} $ & $ \cdots $ & $ N_{J1}-Y_{J1} $&\cellcolor{ProcessBlue} $ N_{11} $ &\cellcolor{ProcessBlue}$ \cdots $ & \cellcolor{ProcessBlue}$ N_{J1} $&\cellcolor{ProcessBlue} $ N_1$  \\
		$ \vdots $  & \cellcolor{Goldenrod}$ \vdots $&  \cellcolor{Goldenrod}$ \ddots $&\cellcolor{Goldenrod}$ \vdots $ &  $ \vdots $&  $ \ddots $&$ \vdots $ &\cellcolor{ProcessBlue}  $ \vdots $& \cellcolor{ProcessBlue} $ \ddots $&\cellcolor{ProcessBlue}$ \vdots $ & \cellcolor{ProcessBlue}$ \vdots $   \\
		$ D $ &\cellcolor{Goldenrod} $ Y_{1D} $ & \cellcolor{Goldenrod} $ \ldots $&\cellcolor{Goldenrod}$ Y_{JD} $ & $ N_{1D}-Y_{1D} $ &  $ \ldots $&$ N_{JD}-Y_{JD} $& \cellcolor{ProcessBlue} $ N_{1D} $ &\cellcolor{ProcessBlue}  $ \ldots $&\cellcolor{ProcessBlue}$ N_{JD} $ &\cellcolor{ProcessBlue} $ N_{D} $\\
		\cmidrule(r){2-4} \cmidrule(lr){5-7} \cmidrule(l){8-10} \cmidrule(l){11-11} 
		Total & $ Y_{1\cdot} $& $ \cdots $ & $ Y_{J\cdot} $&$ N_{1\cdot}-Y_{J\cdot} $ & $ \cdots $ & $ N_{1\cdot}-Y_{J\cdot} $&\cellcolor{ProcessBlue}$ N_{1\cdot}$ &\cellcolor{ProcessBlue}$  \cdots $ &\cellcolor{ProcessBlue} $ N_{J\cdot} $&\cellcolor{ProcessBlue} $ N $ \\
		\bottomrule
	\end{tabular}
	
	\begin{quote}\sffamily \flushleft
		\textit{Legend}
		
		\crule[ProcessBlue]{.3cm}{.3cm} assumed to be known in case study
		
		\crule[Goldenrod]{.3cm}{.3cm} quality standard enforced in case study
	\end{quote}
\end{table}

In the present case study, the Office for National Statistics' census quality threshold is applied to  tabular cells  $ Y_{jd} $ (headcounts of individuals reporting an \ac{LLTI}) but not to cells $ N_{jd} - Y_{jd} $ (headcounts of individuals reporting no \ac{LLTI}). Although the Office for National Statistics' quality standard would normally apply to these cells as well, we consider that most public health statistics producers would only publish prevalence or disease headcounts estimates, not estimates of disease-free headcounts. For simplicity, we neither predict nor test the precision of \ac{LLTI}-free cells. Furthermore, we set the tolerable cell suppression rate to $ \krul = 0 $. Because we aim for no suppression at all, there are no longer any benefits in weighting the loss function (\ref{p3:lossfunction}) by the underlying population sizes, as discussed in section \ref{p3:sec:methods:criteria}. In those conditions, we substitute the loss function used at line \ref{p3:algo1:lossfunction} of Algorithm \ref{p3:algo1} with the simpler crude proportion of suppression below:
\begin{equation}\label{p3:casestudylossfunction}
\ell(f_k)_l \leftarrow  \sum_{j=1}^{J} \sum_{d=1}^{D} \left(     \ I\left[\widehat{\text{RSE}}_{jdl} > 0.2\right] \right)
\end{equation}

Since the quality threshold does not apply to all cross-classifications, the above loss $ \ell $ is only computed over cross-classifications $ jd $ such that $ Y_{jd}/N_{d} \geq0.03 $. Because $ Y_{jd} $ is in reality unknown, we compute the estimated loss $ \hat{\ell} $ over cross-classifications $ jd $ such that $ \theta_{jd}/N_{d} \geq0.03 $. We nevertheless report both the true and the estimated loss in results in order to assess any potential divergence in the algorithm's progression.

Using data from the 2011 UK census table LC3101EWLS \citep{ONScensusLC3101}, 18 age groups were collapsed in order to bring $ Y_{jd}/N_{d} $ to an average level close to 3 per cent or more. With $ D=1,956 $ and setting $ J=6 $, we have a total $JD=11,736$ cells, 49 per cent of which are eligible for the quality standard of a 20 per cent \ac{RSE}. Other settings are configured at $ f_a = 0.01 $, $ f_b=0.04 $, $ h=0.01 $ and $ L=100 $. The acceptable risk of exceeding the tolerable rate of suppression $ \krul $ is set to $ \gamma = 0.01 $.

\subsection{Model specification} \label{p3:sec:models} 
We consider the below working model (\ref{p3:eq:m.general}). Sample counts $ y_{jd} $ are treated as the realisation of a binomial distribution with sample size $ n_{jd} $ and success parameter $p_{jd}=\text{logit}^{-1}(\theta_{jd})$. 
\begin{equation}
\begin{aligned} \label{p3:eq:m.general}
& y_{jd}  \sim \text{Binomial}\left(\text{logit}^{-1}(\theta_{jd}) , n_{jd} \right)      \\
&  \theta_{jd} = \mathbf{\beta}^{(1)}_{j} +  X_d\mathbf{\beta}^{(2)}_{j} +  \upsilon_{d}  + \nu_{jd}       \\
\end{aligned}
\end{equation}
where $ \beta^{(1)}  $ is an $ i $-dimensional vector of age contrasts (with $ \beta^{(1)}_1 = 0 $ for identifiability); $ \mathbf{X} $ a matrix of area-level scaled covariates; $ \mathbf{\beta}^{(2)} $ is a matrix of coefficients controlling the effect of one standard deviation in the covariates on area-level log-odds of \ac{LLTI}; a structured \ac{ICAR} area random effect $\upsilon_{d} $; an unstructured (exchangeable) area by age random effect $  \nu_{jd} $.
 
Covariates in $ \mathbf{X} $ are taken from public sources; namely: the \ac{MSOA}- and district-level \ac{ISAR} \citep{PHEemergencyadm};  the 2015 Income Deprivation Affecting Children Index (proportion of all children aged 0--15 years living in income-deprived families based on tax and benefit departmental database, \citealp{PHE.IDACI2015}); the \ac{MSOA} mean price of residential property sales in 2011 \citep{HPSSA2015}; 2011 mortality ratios indirectly standardised by sex and age \citep{ONSmortality}; and the 2011 \ac{MSOA} Rural Urban Classification consisting of five contrasts \citep{ONSRUC2011}.
No age-specific area covariates were available, which reduces the predictive capabilities of the model: although \acp{ISAR} exhibit strong levels of correlation with crude \ac{LLTI} prevalence across \acp{MSOA}, associations with age-specific prevalence are weaker.

The structured \ac{ICAR} area effect $  \upsilon_{d} $ is included to model the shared spatial surface in disease prevalence as first introduced by \citet{Besag1974} under a sum-to-zero constraint for identifiability: $  \sum_{d=1}^{D} \upsilon_d = 0$. Every structured area effect  $ \upsilon_d  $ is dependent on other area structured effects $\bm{\upsilon}_{[d]} = \{ \upsilon_j, j \neq d \}$  under the following conditional distribution: 
\begin{equation}
\upsilon_d \ |\  \bm{\upsilon}_{[d]} \sim \text{Normal}\left(\sum_{d \neq j} \dfrac{w_{dj} \upsilon_{j}}{w_{dj}} , \dfrac{1}{\tau_\upsilon \sum_{[d]} w_{dj}}\right)
\end{equation}

where spatial weights $ w $ are taken from the spatial dependence matrix $ \mathbf{R}_\upsilon^{-1} $ defined as a $ D \times D $-dimensional contiguity matrix (Queen's method) as below:
\begin{equation}
w_{ij}  =   \begin{cases}
1  & \qquad i = j    \\
1 &  \qquad i, j \ \text{are neighbours}     \\
0  & \qquad \text{otherwise}  
\end{cases}
\end{equation}

corresponding with the below joint prior density:
\begin{equation}
\pi(\bm{\upsilon} \ |\ \tau_\upsilon)  \ \propto \   \exp  \left(-\dfrac{\tau_\upsilon}{2} \ \bm{\upsilon}' \mathbf{R}_\upsilon  \bm{\upsilon} \right)  
\end{equation}

$\bm{\nu}  $ has a basic normal exchangeable prior centred around zero with a unique precision $ \tau_{\nu} $. The sum of $ \bm{\upsilon} $ and $ \bm{\nu} $ forms the widely used convolution prior \citep{Besag1991}. A discrete mixture of normal exchangeable effects with unequal variances was considered to take into account evidence of heteroskedasticity across age groups. This has not led to substantial improvement and has thus been abandoned.

Models are estimated using \ac{INLA} \citep{Rue2009b}, given the substantial computational gains over Markov Chains Monte Carlo. Simulation results for generalised linear mixed models \citep{Carroll2015, Grilli2014b} have shown that posterior distributions obtained are virtually perfectly aligned to posteriors obtained with Markov Chains Monte Carlo sampling. Estimation is implemented using software packages \texttt{INLA} v 1.698, \texttt{R} 3.2.1, and package \texttt{R-INLA} \citep{Martins2013,RCoreTeam2014a} on high performance computer IRIDIS 4, using 16-core nodes, each equipped with 2.6 GHz CPUs and 4 GB RAM.

\begin{figure}\centering
	\includegraphics[width=8cm]{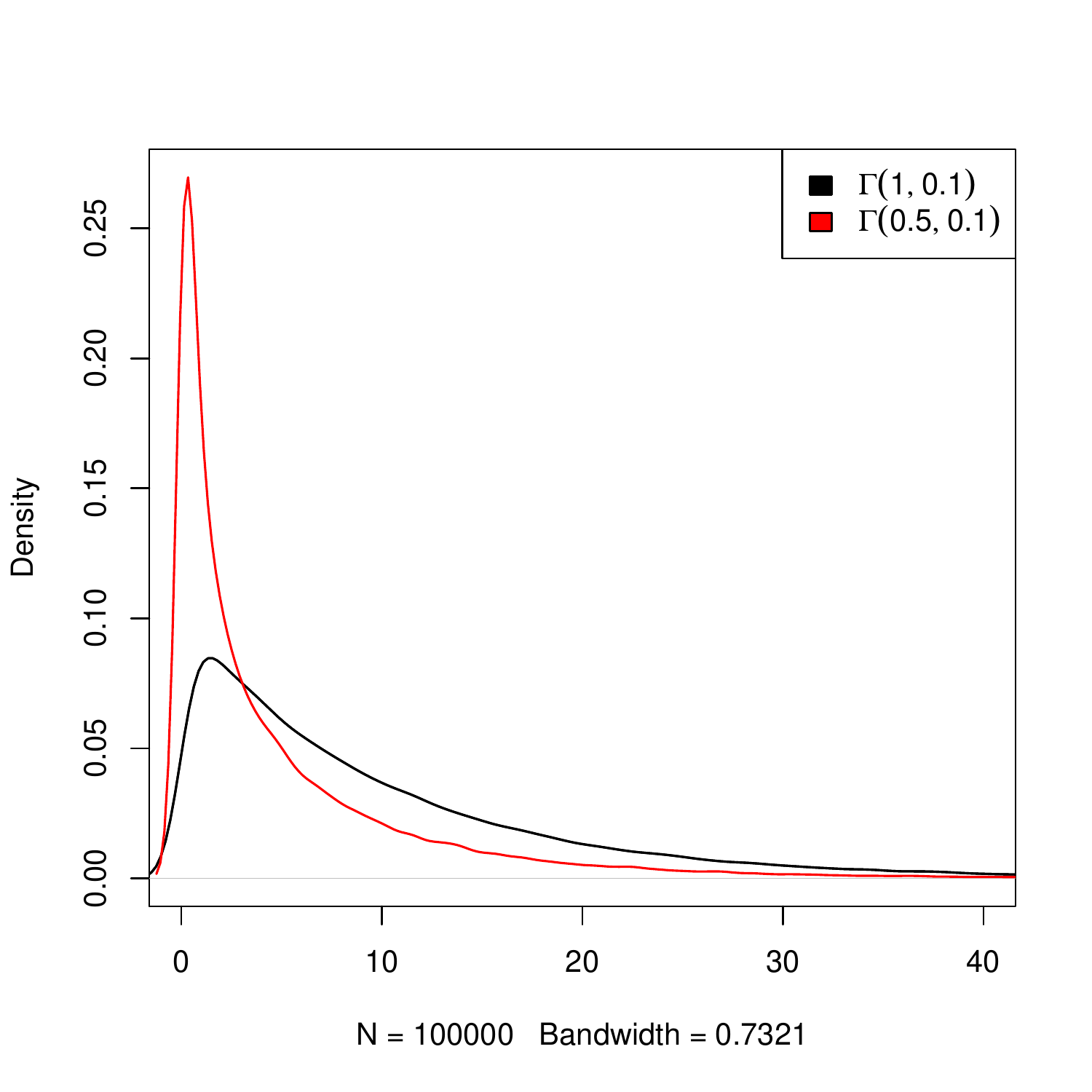}
	\caption{Probability density function of area effects precision hyperpriors} \label{p3:fig:hyperpriors}
\end{figure}
\subsection{Study and fitting priors}
The fitting prior can be formed based on the accumulation of modelling experience on the outcome of interest, particularly around the scale of random effects. It can be difficult to anticipate the magnitude of residual between-area heterogeneity once covariates are used, but the accumulation of evidence over  time can guide priors around the magnitude of area random effects expressed as log-odds. We specify inverse Gamma priors for these effects on the basis of previous evidence that once covariates are introduced, their variance is likely to be near 0.1,
and that half of this between-area heterogeneity is likely to be spatially structured. Following this, we place equal weight on structured and unstructured effect. To achieve a sum variance of 0.1, we need a mean hyperprior precision of 5 for each area effect. Based on plots we also decide that the variance of this hyperprior should be ten times the mean to avoid dominating the data. This leads to hyperpriors $ \Gamma(0.5, 0.1) $ being placed on each of the inverse variance of area effects or 
$ \Gamma(1,  0.1) $ when only unstructured effects are used (see Figure \ref{p3:fig:hyperpriors}).

In contrast, design priors are obtained from the marginal posterior density of $ \bm{\theta} $ obtained from fitting $ \mathcal{M} $ on a simulated pilot sample $ s_0 $ with sampling fraction $ f=0.01 $. Where prior knowledge expressed in fitting priors is very informative, the sampling fraction can be relatively small and still produce relatively narrow design priors.

\subsection{Design-based simulation} \label{p3:sec:simulationdesign}
A second series of simulations is carried out with a design-based procedure: samples are taken directly from the census tables using a multinomial law. Unlike with the model-based simulation, results no longer assume that the models are true. Doing so makes it possible to verify the validity of the sample size obtained under the previously described model-based \ac{SSD} procedure. We also take the opportunity to estimate the relative efficiency in a series of scenarios of ascending complexity, from the simple design-based estimator to an \ac{HB} predictor borrowing strength from a covariate and an explicit spatial covariance structure. These scenarios notated  S1 to S4 are summarised in Table \ref{p3:tab:DBsimscenarios}. To this end, we simulate sampling and estimating each one and compare it with the true population value.

\begin{table}[t]
	\caption{Description of design-based simulation scenarios} \label{p3:tab:DBsimscenarios}
	\centering    
	\begin{tabular}{lll}
		\toprule
		Scenario & Estimation method &  Model structure for predictor $ \theta_{jd} $    \\
		\midrule
		S1 & Direct estimation & --     \\
		S2 & \ac{HB} -- no covariate  & $  \bm{\beta}^{(1)}_{j}  +\nu_{jd} $      \\ 
		S3 & \ac{HB} -- covariate  &$  \bm{\beta}^{(1)}_{j} + X_d\bm{\beta}^{(2)}_d +    \nu_{jd}  $      \\  
		S4 & \ac{HB} -- covariate + spatial structure  & $  \beta^{(1)}_{j} +  X_d\bm{\beta}^{(2)}_{j} +  \upsilon_{d}  + \nu_{jd}  $      \\
		\bottomrule
	\end{tabular} \medskip
	\begin{quote} \sffamily \footnotesize
		\textit{Note:} HB: Hierarchical Bayes; MSOA: Middle Layer Super Output Area. 
	\end{quote}
\end{table}

One hundred simple random samples $ s_b $ are drawn from $ U $  with simulated sizes  $  n_{jdb} \sim \text{Binomial}(N_{jdb}, f) $ where $ f $ is set by the experiment, with no guarantee that all $ n_{jdb} >0 $. We simulate sample headcounts such that:
\begin{equation} 
y_{jdb} \sim \text{Binomial}(n_{jdb}, N_{jdb}^{-1} \ Y_{jdb})
\end{equation}

Samples $ s_b $ are used to produce model-based estimates $ \widehat{\overline{Y}}_{jdb} $ of prevalence $ \overline{Y}_{jdb} =   Y_{jdb}/N_{jdb}  $ of \ac{LLTI} every area making up the population as a function of characteristics of this population.

To reduce the computational burden of this simulation study, we only estimate relative efficiency under two sampling fractions: 2 and 4 per cent. The number of iterations is set to $ B=400 $ for each combination of a scenario and sampling fraction, totalling 2,400 procedures of model estimation and prediction. For scenarios S2--S5, and for each of the chosen sampling fractions $ f $, we compute measures of accuracy:
\begin{itemize}
	\item the \acl{RMSE} (\acs{RMSE}) 
	\begin{equation}
	\text{RMSE}_{jd} = \sqrt{B^{-1} \sum_{b=1}^{B} \left(\widehat{\overline{Y}}_{jdb}- \overline{Y}_{jdb} \right)^2}
	\end{equation}
	\item the bias
	\begin{equation}
	\text{Bias}_{jd} = B^{-1} \sum_{b=1}^{B}  \left(\widehat{\overline{Y}}_{jdb}- \overline{Y}_{jdb}\right)  
	\end{equation}
	\item the absolute relative bias (\ac{ARB})
	\begin{equation}
	\text{ARB}_{jd} = B^{-1} \sum_{b=1}^{B}  \dfrac{|\widehat{\overline{Y}}_{jdb}- \overline{Y}_{jdb}|}{\overline{Y}_{jdb}}
	\end{equation}
	\item the relative \acs{RMSE} or \acl{RSE} (\acs{RSE}) 
	\begin{equation}
	\text{RSE}_{jd} = \dfrac{\text{RMSE}_{jd}}{{\overline{Y}}_{jd}}
	\end{equation}
	\item the \ac{RSE}'s relative bias (RSEB) (where $ \widehat{\text{postMSE}} $ is the posterior variance)
	\begin{equation}
	\begin{aligned}
	& \text{RSEB}_{jd} = B^{-1} \sum_{b=1}^{B} \dfrac{\widehat{\text{RSE}}_{jdb}-  \text{RSE}_{jdb} }{\text{RSE}_{jdb} }     \\
	& \widehat{\text{RSE}}_{jd} = \dfrac{\sqrt{\widehat{\text{postMSE}}_{jd}}}{\hat{\bar{Y}}_{jd}}
	\end{aligned}
	\end{equation}
\end{itemize}

These measures are mapped and averaged across small areas $ d $ to be reported in tables. For scenario S1, the bias of sample means is by definition zero and we calculate the \ac{MSE} and \ac{RSE} using the variance formula of the sample proportion:
\begin{equation} 
\text{Var}(\bar{y}_{jd}) = \dfrac{\overline{Y}(1-\overline{Y})}{n_{jd}} =  \dfrac{\overline{Y}(1-\overline{Y})}{ f \ N_{jd} }
\end{equation}

\section{Results} \label{p3:sec:results}
\subsection{\acs{SSD} procedure}
The basic model (\ref{p3:eq:m.general}) is fitted to pilot sample $ s_0 $ of \ac{ESS} $ n=146,574$, with 1,952 cells $ jd $ containing no observations. This means no data is available for 1.3 per cent of the total number of cells, or 8.8 per cent of cells where the minimum quality standard applies. The fitted model achieves an acceptable \ac{DIC} of 33,787. The \ac{SSD} procedure summarised in Table \ref{p3:tab:MBstudyresults} produces an effective sample fraction requirement of 2.6875 per cent, equivalent to an \ac{ESS} of 394,103. Assuming the procedure is correct, this solution is considered to fall at the most 0.1875 percentage points from the true effective sampling fraction achieving a negligible risk of cell suppression. Further iterations would reduce this distance. The algorithm was guided by an estimator of the loss function $ \ell $ which depends on predictors $ \theta $ to determine which cells should be checked for the quality standard. Any bias on $ \theta $ can potentially increase or decrease the suppression risk to a substantial degree. In this case study, we note that the estimator $ \hat{\ell} $ fails to detect a risk of suppression at step $ k=3 $, which leads the algorithm to select a lower interval of sample fractions that it would have done if it had been driven by actual loss $ \ell $  calculated using the census figures $Y_{jd}$.

\begin{table}
	\centering \footnotesize
	\caption{Case study results of the \acs{SSD} procedure after four steps}\label{p3:tab:MBstudyresults}
	\setlength{\tabcolsep}{4pt}
	\begin{tabular}{rrrrrrr}
		\toprule
		Step &     Sampling &           Mean loss &                 Mean loss &                  Risk &                        Risk &        Search \\
		$ k $ & rate $ f_k $ & $ \bar{\ell}(f_k) $ & $ \hat{\bar{\ell}}(f_k) $ & $ \Pr(\ell(f_k)  >0)$ & $ \Pr(\hat{\ell}(f_k) >0) $ &      interval \\ \midrule
		1 &       0.010000 & 0.004000 & 0.008400 & 0.063500 &  0.095200 &  \\
		1 &       0.040000 & 0.000000 & 0.000000 & 0.000000 &  0.000000 &   \\
		1 &       0.025000 & 0.000006 & 0.000005 & 0.025000 &  0.010000 &     [0.010000,0.040000] \\
		2 &       0.032500 & 0.000000 & 0.000000 & 0.000000 &  0.000000 &   [0.025000,0.040000] \\
		3 &       0.028750 & 0.000003 & 0.000000 & 0.010050 &  0.000000 & [0.025000,0.032500] \\
		4 &       0.026875 & 0.000001 & 0.000000&  0.005076 &  0.000000   & [0.025000,0.028750] \\ \bottomrule
	\end{tabular}\medskip
	
	\flushleft{\footnotesize \sffamily \textit{Note:} Authors' own calculations. Loss and risk estimated over 200 simulations per step.}
\end{table}

Traditional survey sampling methods can determine the most cost-effective strategy to achieve this target \ac{ESS}. A basic household survey like the \ac{EHS} is based on a multistage design sampling one household per dwelling per address selected in the address sampling frame, and interviewing all eligible members (by proxy if necessary). Using the 2010/11 microdata sample for England of 11,188 effective responses, we estimated the \ac{DEFF} to 1.16 for the \ac{LLTI} question. This implies that survey planners should aim for a final sample size at least 16 per cent higher than the \ac{ESS} found above, to account for the design effect. Yet many surveys have additional sampling stages which can increase the \ac{DEFF}, for instance selecting postcode sectors as primary sampling units and then addresses/dwellings/households as secondary sampling units. This is for instance the case of the Health Survey for England for which we estimated a \ac{DEFF} of 1.44 using a microdata sample from 2011 \citep{HSE2011}. The \ac{DEFF} can easily be used in area-level \ac{SAE} models in a variance function, or in unit-level generalised linear mixed models with the \ac{ESS} method described by \citet{Chen2014}.

\subsection{Design-based simulation}
We verify the validity of the small area model used in this case study as well as the reliability of the loss function estimator across iterations of the algorithm. Design-based simulation results are presented in Tables \ref{p3:tab:resultsS1} to \ref{p3:tab:resultsS4}. 

Results provide evidence that the relative posterior variances obtained from the spatial model are inflated by 20 to 40 per cent in basic scenario S2 (Table \ref{p3:tab:resultsS2}) and by as much as 40 to 50 per cent in scenario S3 (Table \ref{p3:tab:resultsS3}). This is likely to make the \ac{SSD} procedure very conservative as many cells will be unnecessarily suppressed. This is a likely sign of poor model specification. Simulations for other scenarios are reported here to quantify the relative efficiency of the different model- and design-based methods. Efficiency savings from using a particularly model or from an increase in sample size can be derived as the ratio of \acs{RMSE} of two scenarios. Comparing the scenarios presented in Table \ref{p3:tab:DBsimscenarios}, we find that even a basic model without area covariates (see S1--S2, Tables \ref{p3:tab:resultsS1} and \ref{p3:tab:resultsS2}) achieves a reduction in \ac{RMSE} by about 49 per cent overall, and up to 74 per cent for youngest age groups ($ f=0.02 $). This comes at a very reasonable computational cost and with a bias of between 0.1 and 0.7 percentage point. With a much larger sample ($ f=0.04 $), the efficiency gain comes down to approximately 35 per cent.

The addition of covariates delivers strong efficiency gains (see S3, Tables \ref{p3:tab:resultsS3}) with an reduction in \ac{RMSE} by 72  ($ f=0.02 $) and 62 per cent ($ f=.04 $). This includes a reduction in bias from 0.02 to below 0.01 percentage point. 

Comparing results for $ f=0.02 $ and $ f=0.04 $, we note that doubling the sample size has diminishing returns as the methods becomes more complicated. Whilst it reduces the standard deviation of design-based estimates S1 by 29 per cent, the \ac{RMSE} of estimates S2 is only reduced by 12 per cent. As for estimates S3 and S4 the reduction by less than 3 and 4 per cent respectively is not material for an augmentation of this magnitude. This shows that \ac{RMSE} as a function of sample size is already relatively horizontal in the regions of sample sizes we are investigating, which is not the case for design-based estimators. Regardless of the sample size, bias remains high. Although the average bias
across areas remains of the order of 0.1 to 0.3 percentage points, the more meaningful \ac{ARB} metric reveals an average absolute bias of 5 to 20 per cent of the target parameter. This represents almost all the total \ac{RSE}. This bias is not very sensitive to the sampling fraction or the type of between-area variance structure used.

\begin{figure}
	\begin{minipage}[c]{.85\linewidth}
		\begin{minipage}[c]{.5\linewidth}
			\includegraphics[width=\linewidth]{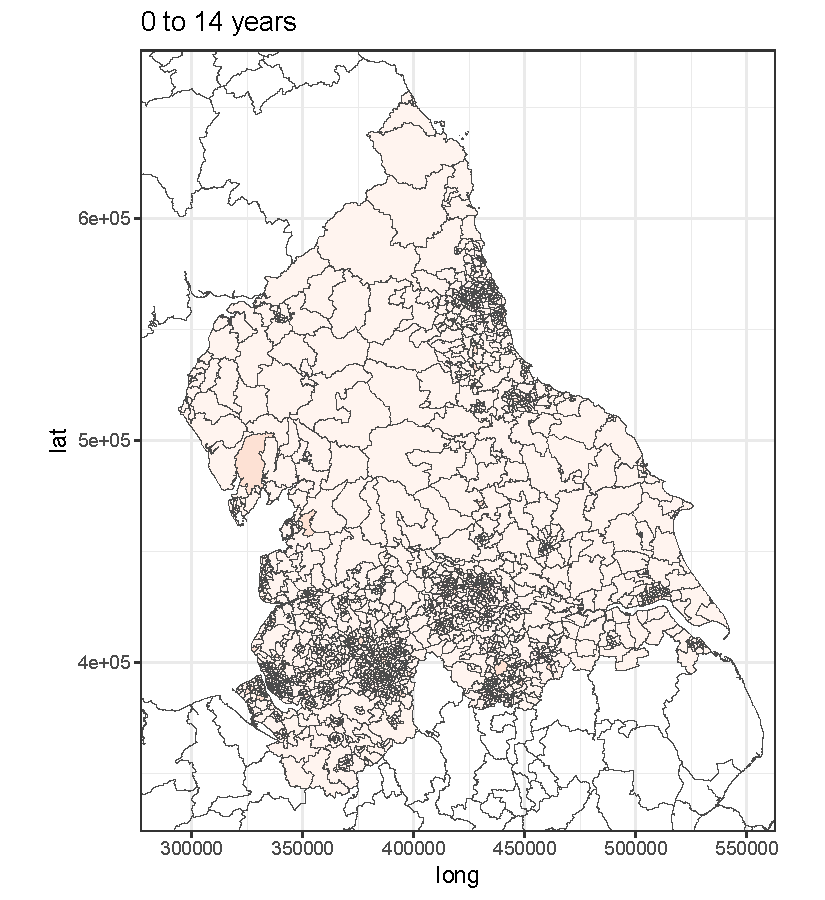}
		\end{minipage}%
		\begin{minipage}[c]{.5\linewidth}
			\includegraphics[width=\linewidth]{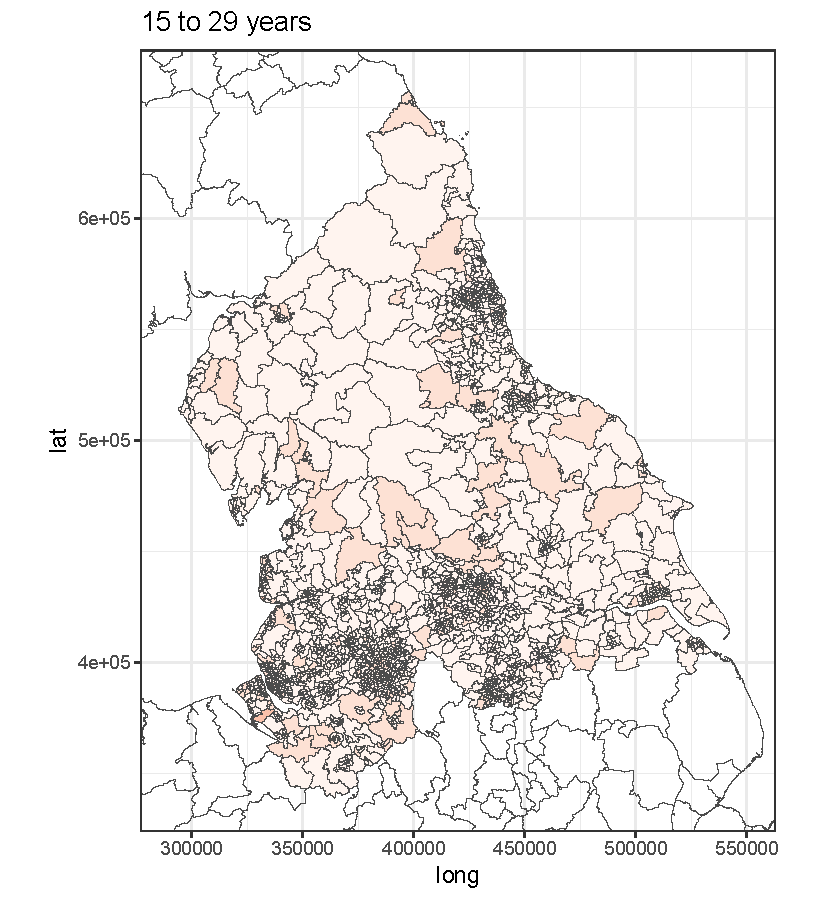}
		\end{minipage}
		
		\begin{minipage}[c]{.5\linewidth}
			\includegraphics[width=\linewidth]{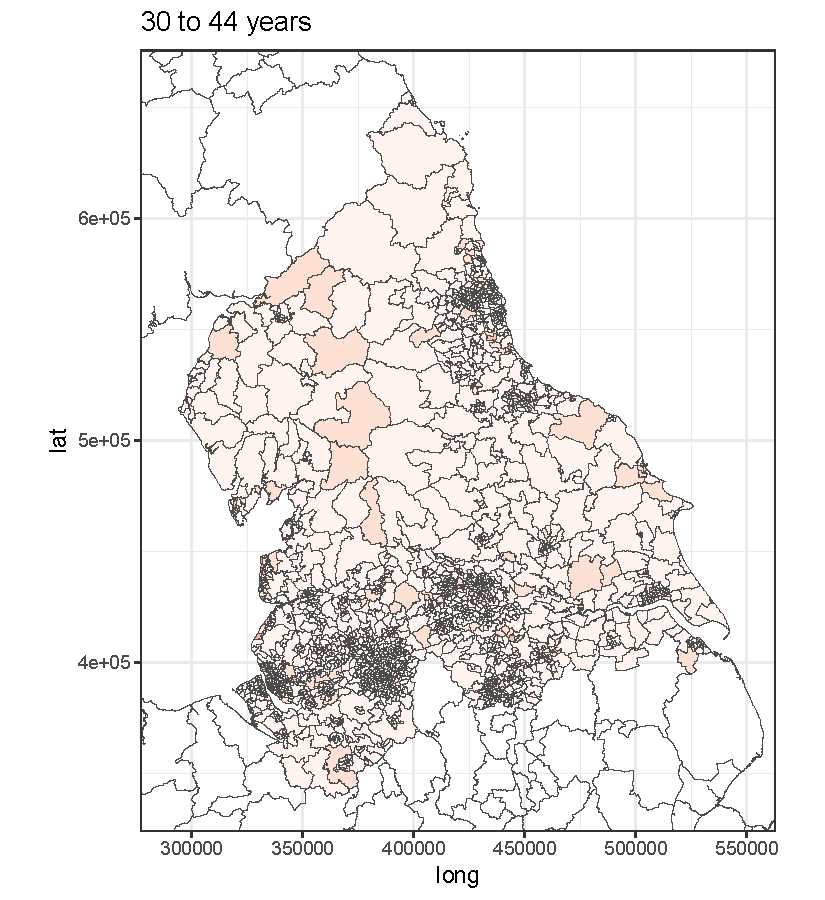}
		\end{minipage}%
		\begin{minipage}[c]{.5\linewidth}
			\includegraphics[width=\linewidth]{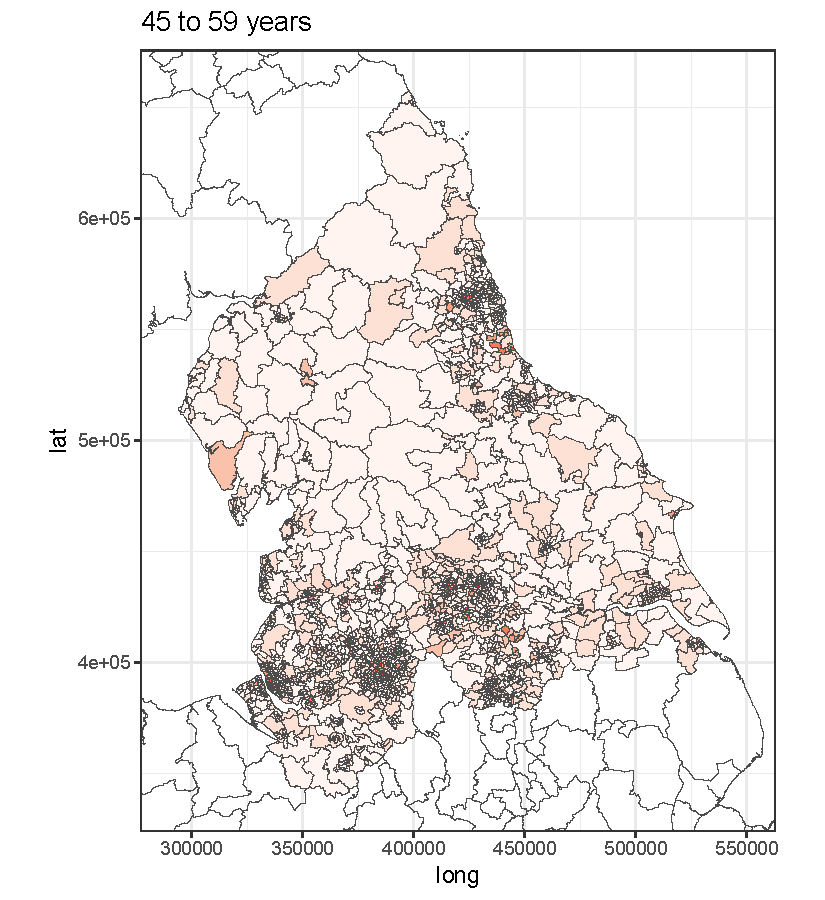}
		\end{minipage}
		
		\begin{minipage}[c]{.5\linewidth}
			\includegraphics[width=\linewidth]{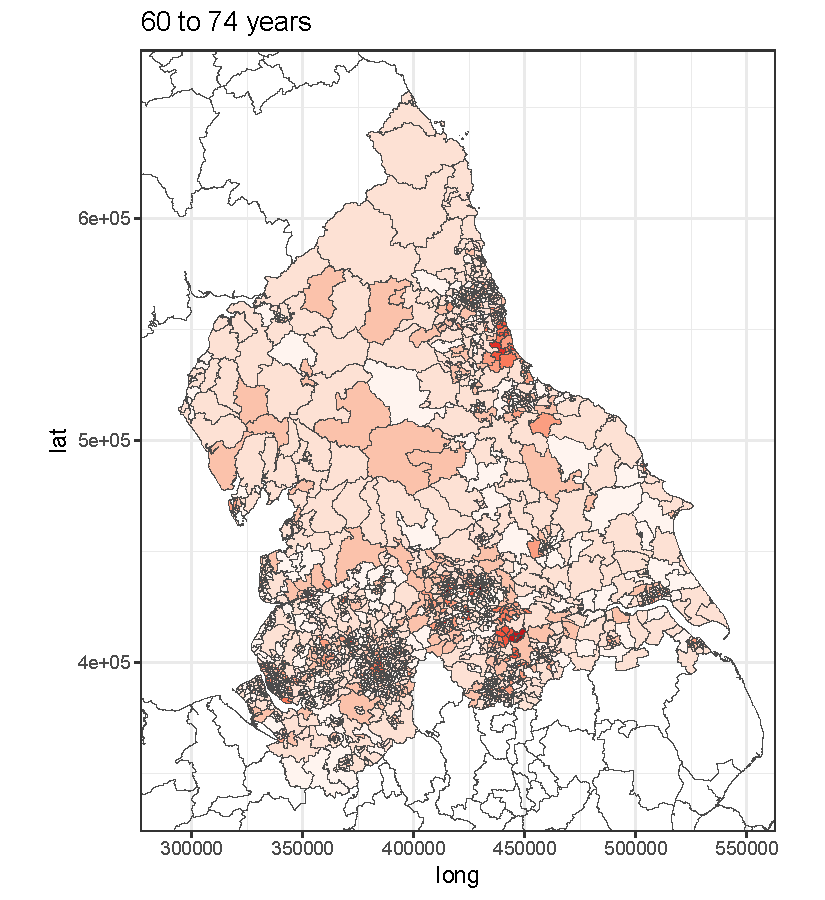}
		\end{minipage}%
		\begin{minipage}[c]{.5\linewidth}
			\includegraphics[width=\linewidth]{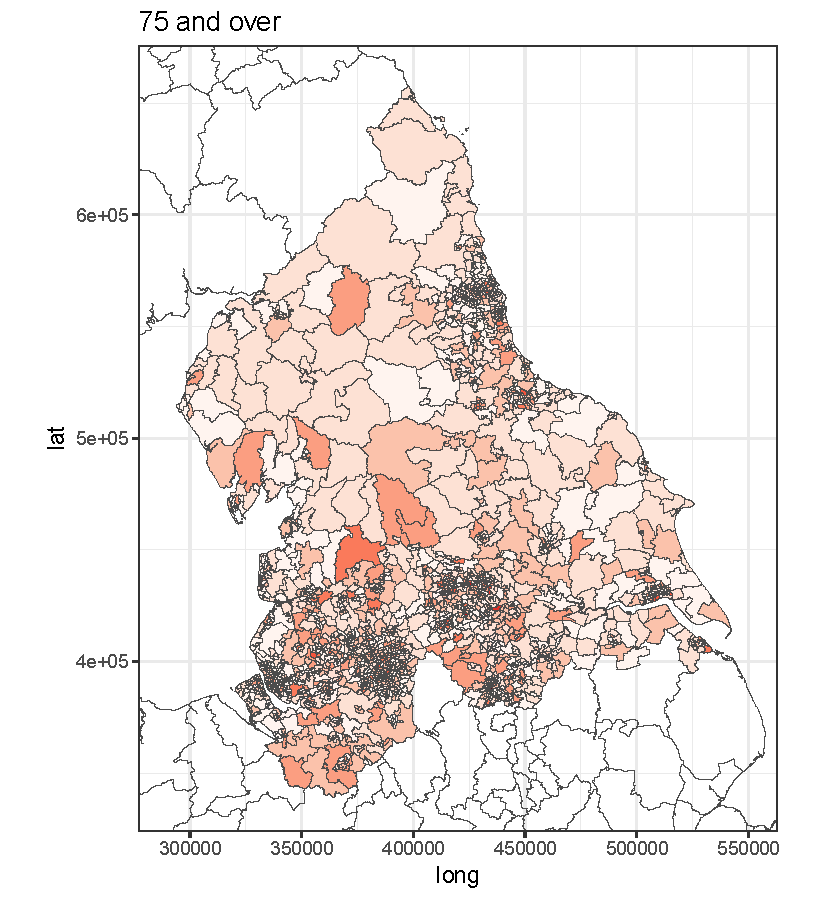}
	\end{minipage}\end{minipage}\begin{minipage}[c]{.15\linewidth}
		\includegraphics[width=1.5cm]{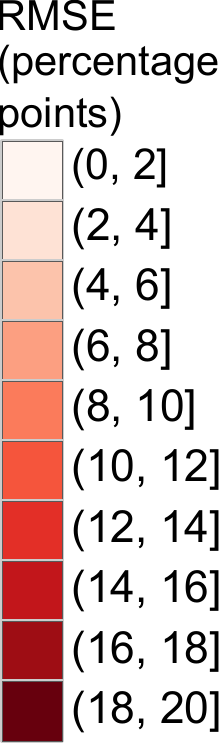}
	\end{minipage}
	
	\caption{Map of age-specific root \acs{MSE} in prediction on the proportion scale across 400 iterations of scenario S3 with sampling fraction $ f=0.02 $} \label{p3:fig:mapRMSE.S3}
\end{figure}

\begin{figure}
	\begin{minipage}[c]{.85\linewidth}
		\begin{minipage}[c]{.5\linewidth}
			\includegraphics[width=\linewidth]{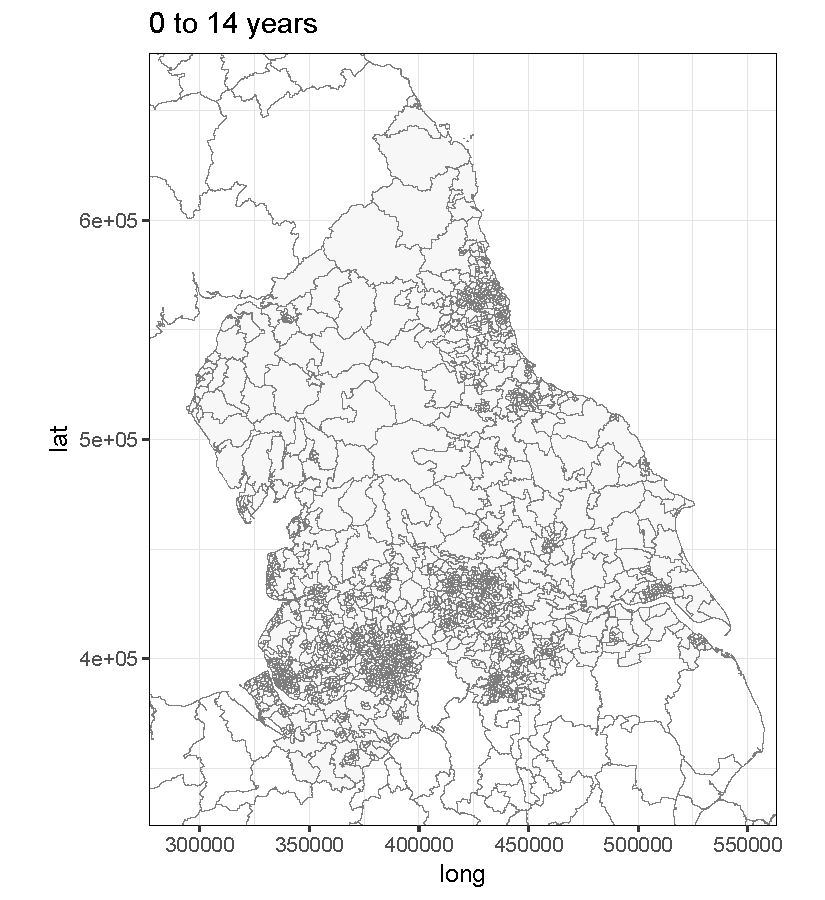}
		\end{minipage}%
		\begin{minipage}[c]{.5\linewidth}
			\includegraphics[width=\linewidth]{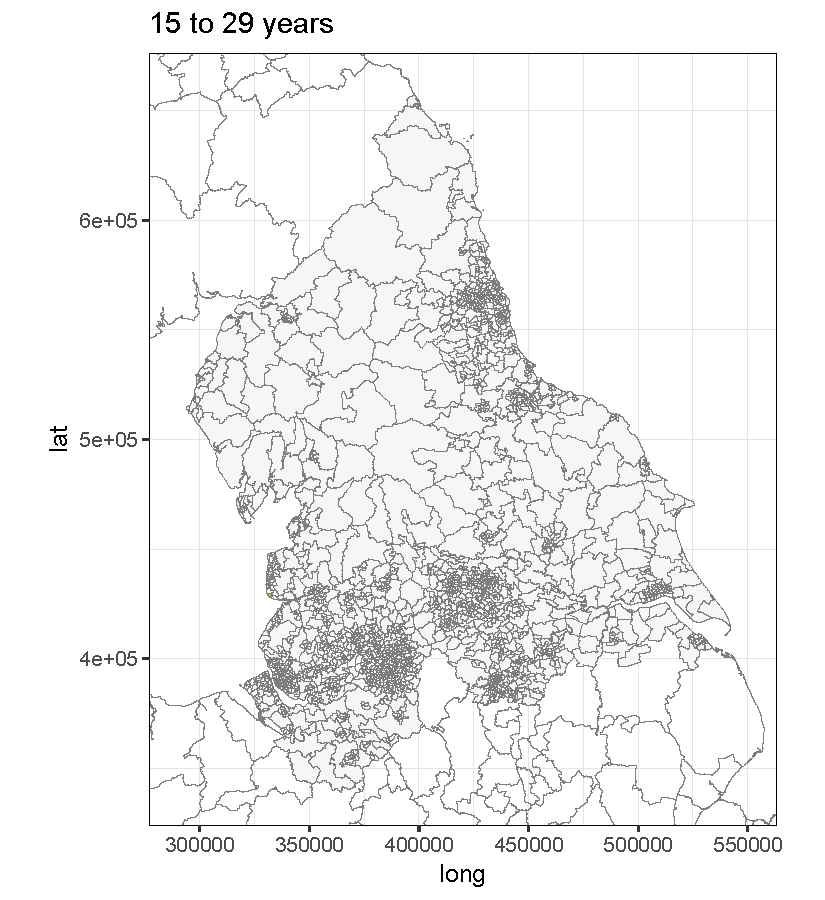}
		\end{minipage}
		
		\begin{minipage}[c]{.5\linewidth}
			\includegraphics[width=\linewidth]{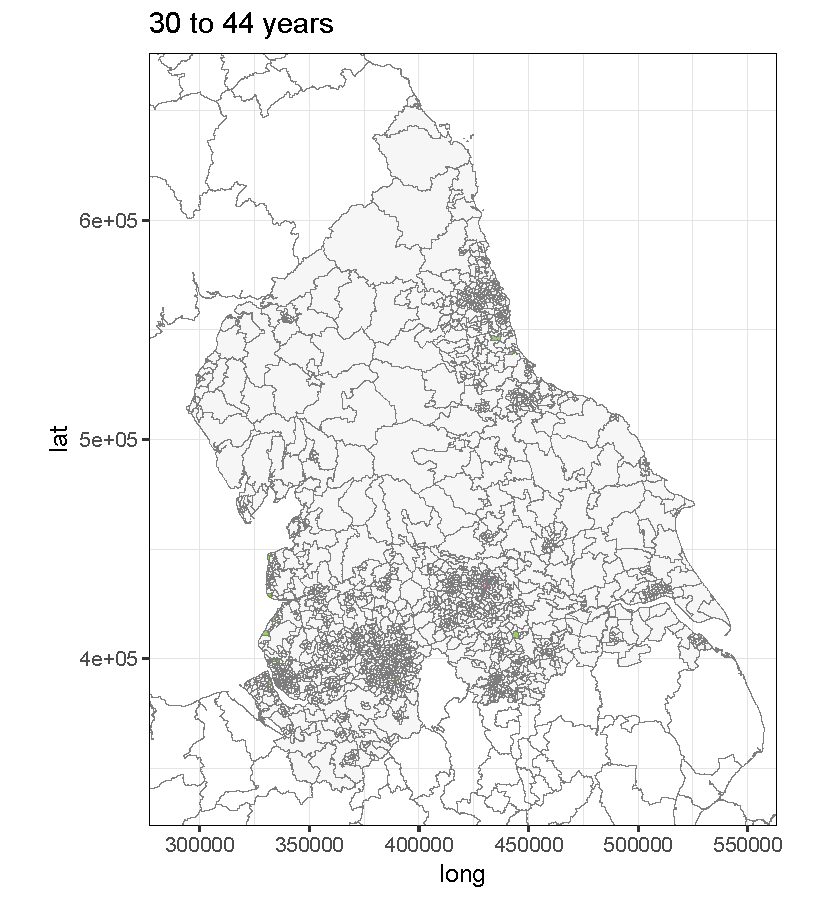}
		\end{minipage}%
		\begin{minipage}[c]{.5\linewidth}
			\includegraphics[width=\linewidth]{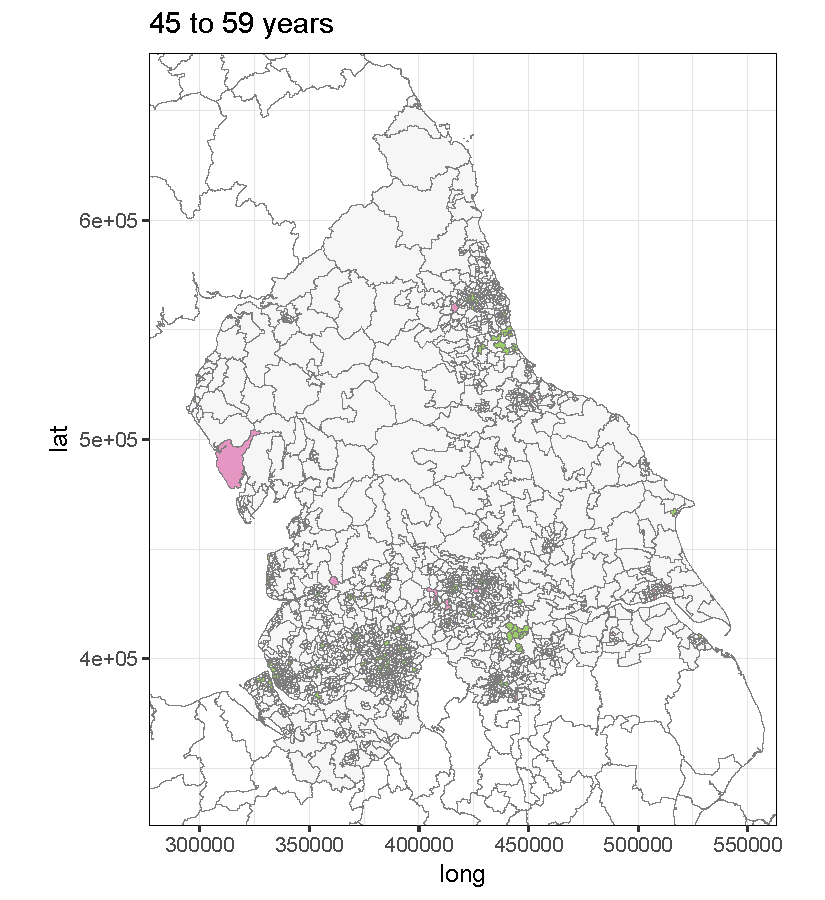}
		\end{minipage}
		
		\begin{minipage}[c]{.5\linewidth}
			\includegraphics[width=\linewidth]{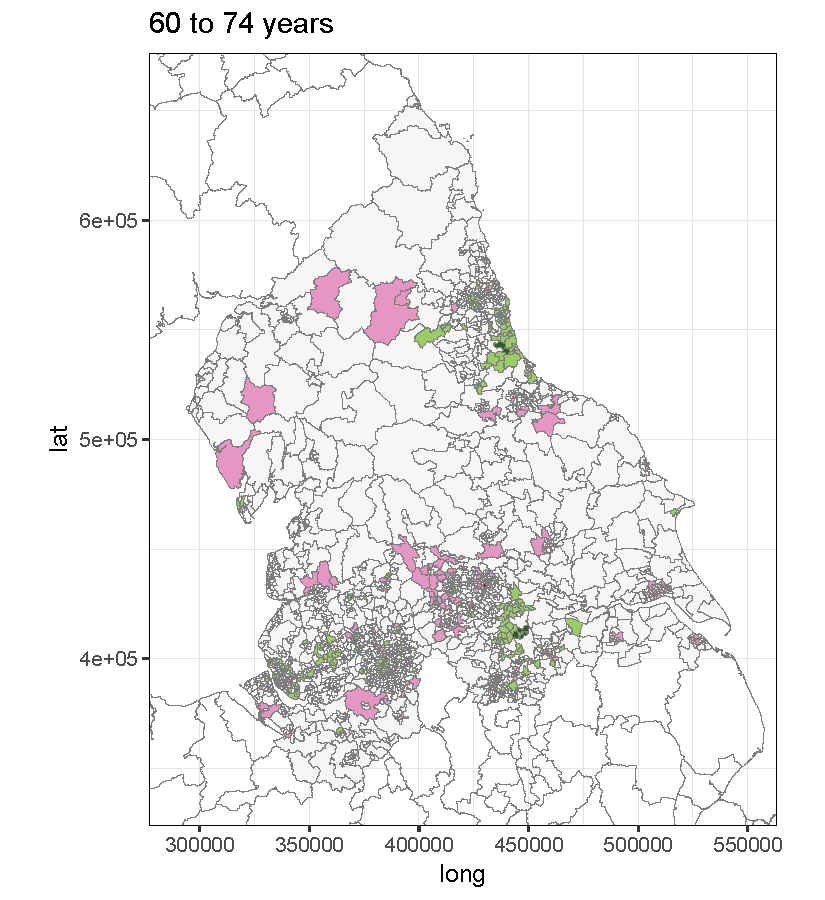}
		\end{minipage}%
		\begin{minipage}[c]{.5\linewidth}
			\includegraphics[width=\linewidth]{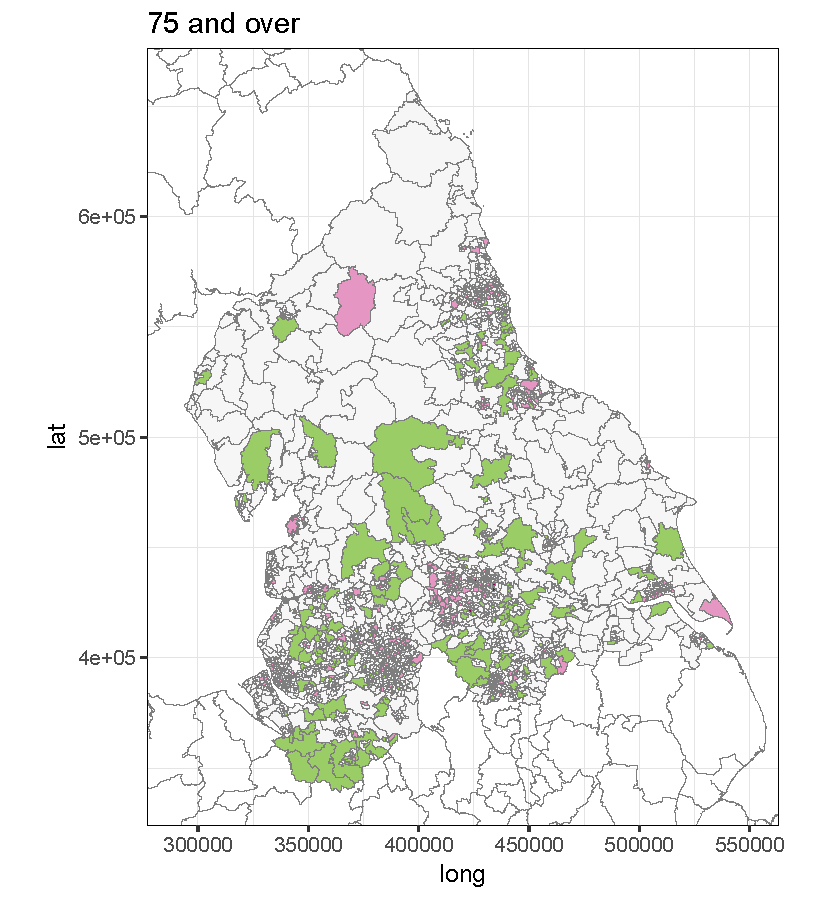}
	\end{minipage}\end{minipage}\begin{minipage}[c]{.3\linewidth}
		\includegraphics[width=2cm]{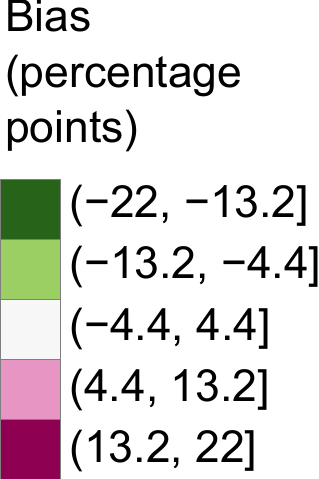}
	\end{minipage}
	\caption{Map of age-specific prediction bias in percentage points across 400 iterations of scenario S3 with sampling fraction $ f=0.02 $} \label{p3:fig:mapbias.S3}
\end{figure}

The working model was designed to borrow strength by assuming that areas situated near each other were more similar. We thus examined simulation results to verify whether this assumption holds everywhere equally. Figures \ref{p3:fig:mapRMSE.S3} and \ref{p3:fig:mapbias.S3} present the \ac{RMSE} and bias for age-specific \ac{MSOA} prevalence proportion computed from S3 with $ f=0.020 $. Hotpots of high \ac{RMSE} and bias of 1 to 2 percentage points are visible which exhibit a spatial pattern. This could signal outlier areas, or areas in which that covariates are poorly measured and making the linear predictor particularly biased. The convolution prior used here may not fully and authentically reproduce the spatial pattern of the health status under consideration.

\begin{table}
	\caption{\label{p3:tab:resultsS1} Simulation S1: mean accuracy metrics by age group}
	\centering  \footnotesize
	\begin{tabular}{lrrrrr}
		\toprule
		\multirow{2}{0.5cm}{Age (years)}   & \multicolumn{5}{c}{\textbf{S1} \textit{(`design-based')}}   \\ 
		\cmidrule(lr){2-6}
		&$ \overline{\text{RMSE}} $& $ \overline{\text{Bias}} $ & $ \overline{\text{RSE}}  $& $ \overline{\text{RSEB}}  $& $ \ell(f)$ \\
		\midrule
		\multicolumn{6}{c}{Sampling fraction $ f = 0.02 $ } \\
		\midrule
		0--14 & 0.0379 & --- & 1.0600 & --- & 0.0000 \\
		15--29 & 0.0449 & --- & 0.8048 & --- & 0.0005 \\
		30--44 & 0.0576 & --- & 0.5663 & --- & 0.1922 \\
		45--59 & 0.0752 & --- & 0.3722 & --- & 0.8226 \\
		60--74 & 0.1050 & --- & 0.2660 & --- & 0.8502 \\
		75+ & 0.1436 & --- & 0.2054 & --- & 0.4054 \\
		All ages & 0.0774 & --- & 0.5458 & --- & 0.3785 \\
		\midrule
		\multicolumn{6}{c}{Sampling fraction $ f = 0.04 $ }  \\
		\midrule
		0--14 & 0.0268 & --- & 0.7495 & --- & 0.0000 \\
		15--29 & 0.0318 & --- & 0.5691 & --- & 0.0005 \\
		30--44 & 0.0407 & --- & 0.4004 & --- & 0.1907 \\
		45--59 & 0.0532 & --- & 0.2632 & --- & 0.6687 \\
		60--74 & 0.0742 & --- & 0.1881 & --- & 0.3129 \\
		75+ & 0.1015 & --- & 0.1452 & --- & 0.0133 \\
		All ages & 0.0547 & --- & 0.3859 & --- & 0.1977 \\
		\bottomrule
	\end{tabular}
	
	\flushleft{\footnotesize \sffamily \textit{Note:} Accuracy metrics are aggregated into a mean across all 1,956 areas. }
\end{table}

\begin{table}\caption{Simulation S2: mean accuracy metrics by age group  across $ B=400 $ replications}\label{p3:tab:resultsS2}
	\footnotesize \centering  \setlength{\tabcolsep}{5.2pt} 
	\begin{tabular}{lrrrrrrrrr} 
		\toprule
		\multirow{1}{0.5cm}{Age (years)}   & \multicolumn{9}{c}{\textbf{S2} \textit{(`HB - no covariates')}}     \\ 
		\cmidrule(lr){2-10} 
		&$ \overline{\text{RMSE}} $& $ \overline{\text{Bias}} $ &$ \overline{\text{ARB}} $ & $ \overline{\text{RSE}}  $& $ \overline{\text{RSEB}}  $& $ \ell(f)$ & $ \overline{\Pr(\ell >0)} $ & $ \hat{\ell}(f) $ & $  \overline{\Pr(\hat{\ell} >0)} $ \\
		\midrule
		\multicolumn{10}{c}{Sampling fraction $ f = 0.02 $ }  \\
		\midrule
		0--14 & 0.0100 & 0.0010 & 0.2648 & 0.3004 & 0.6734 & 0.0000 & 0.000 & 0.0000 & 0.000 \\
		15--29 & 0.0149 & 0.0010 & 0.2340 & 0.2750 & 0.5234 & 0.0005 & 1.000 & 0.0057 & 1.000 \\
		30--44 & 0.0316 & 0.0021 & 0.2876 & 0.3298 & 0.1488 & 0.1921 & 1.000 & 0.1027 & 1.000 \\
		45--59 & 0.0527 & -0.0017 & 0.2317 & 0.2710 & 0.0418 & 0.7449 & 1.000 & 0.8589 & 1.000 \\
		60--74 & 0.0722 & -0.0071 & 0.1576 & 0.1859 & 0.0780 & 0.1670 & 1.000 & 0.1649 & 1.000 \\
		75+ & 0.0568 & -0.0069 & 0.0691 & 0.0814 & 0.4325 & 0.0000 & 0.000 & 0.0000 & 0.000 \\
		All ages & 0.0397 & -0.0019 & 0.2075 & 0.2406 & 0.3163 & 0.1841 & 1.000 & 0.1887 & 1.000 \\
		\midrule
		\multicolumn{10}{c}{Sampling fraction $ f = 0.04 $ }  \\
		\midrule
		0--14 & 0.0100 & 0.0010 & 0.2547 & 0.2984 & 0.4418 & 0.0000 & 0.000 & 0.0000 & 0.000 \\
		15--29 & 0.0146 & 0.0012 & 0.2211 & 0.2681 & 0.3228 & 0.0005 & 1.000 & 0.0005 & 1.000 \\
		30--44 & 0.0280 & 0.0021 & 0.2442 & 0.2893 & 0.0609 & 0.1727 & 1.000 & 0.1727 & 1.000 \\
		45--59 & 0.0437 & -0.0006 & 0.1849 & 0.2229 & -0.0075 & 0.3322 & 1.000 & 0.3322 & 1.000 \\
		60--74 & 0.0600 & -0.0047 & 0.1268 & 0.1539 & 0.0220 & 0.0107 & 1.000 & 0.0107 & 1.000 \\
		75+ & 0.0539 & -0.0062 & 0.0638 & 0.0773 & 0.2697 & 0.0000 & 0.000 & 0.0000 & 0.000 \\
		All ages & 0.0351 & -0.0012 & 0.1826 & 0.2183 & 0.1849 & 0.0860 & 1.000 & 0.0879 & 1.000 \\
		\bottomrule
	\end{tabular}
	
	\flushleft{\footnotesize \sffamily \textit{Note:} Accuracy metrics are aggregated into a mean across all 1,956 areas. }
\end{table}

\begin{table}\caption{Simulation S3: mean accuracy metrics by age group  across $ B=400 $ replications} \label{p3:tab:resultsS3}
	\centering  \footnotesize \setlength{\tabcolsep}{5.2pt} 
	\begin{tabular}{lrrrrrrrrr}
		\toprule
		\multirow{2}{0.5cm}{Age (years)}   & \multicolumn{9}{c}{\textbf{S3} \textit{(`HB -- with covariates')}}    \\ 
		\cmidrule(lr){2-10} 
		&$ \overline{\text{RMSE}} $& $ \overline{\text{Bias}} $ &$ \overline{\text{ARB}} $ & $ \overline{\text{RSE}}  $& $ \overline{\text{RSEB}}  $& $ \ell(f)$ & $ \overline{\Pr(\ell >0)} $ & $ \hat{\ell}(f) $ & $  \overline{\Pr(\hat{\ell} >0)} $ \\
		\midrule
		\multicolumn{10}{c}{Sampling fraction $ f = 0.02 $ }  \\
		\midrule
		0--14 & 0.0078 & -0.0009 & 0.2073 & 0.2134 & 0.4473 & 0.0000 & 0.000 & 0.0000 & 0.000 \\
		15--29 & 0.0108 & -0.0026 & 0.1857 & 0.1929 & 0.4871 & 0.0000 & 0.000 & 0.0000 & 0.008 \\
		30--44 & 0.0143 & -0.0006 & 0.1269 & 0.1371 & 0.6470 & 0.0000 & 0.000 & 0.0000 & 0.000 \\
		45--59 & 0.0255 & -0.0016 & 0.1107 & 0.1222 & 0.4717 & 0.0000 & 0.003 & 0.0000 & 0.003 \\
		60--74 & 0.0360 & -0.0020 & 0.0814 & 0.0903 & 0.4439 & 0.0000 & 0.000 & 0.0000 & 0.000 \\
		75+ & 0.0346 & 0.0024 & 0.0467 & 0.0496 & 0.5439 & 0.0000 & 0.000 & 0.0000 & 0.000 \\
		All ages & 0.0215 & -0.0009 & 0.1264 & 0.1343 & 0.5068 & 0.0000 & 0.003 & 0.0000 & 0.008 \\
		\midrule
		\multicolumn{10}{c}{Sampling fraction $ f = 0.04 $ }  \\
		\midrule
		0--14 & 0.0076 & -0.0008 & 0.2004 & 0.2080 & 0.3958 & 0.0000 & 0.000 & 0.0000 & 0.000 \\
		15--29 & 0.0104 & -0.0024 & 0.1769 & 0.1865 & 0.3872 & 0.0000 & 0.000 & 0.0000 & 0.000 \\
		30--44 & 0.0141 & -0.0003 & 0.1223 & 0.1359 & 0.4695 & 0.0000 & 0.000 & 0.0000 & 0.000 \\
		45--59 & 0.0248 & -0.0014 & 0.1044 & 0.1190 & 0.3044 & 0.0000 & 0.000 & 0.0000 & 0.000 \\
		60--74 & 0.0348 & -0.0016 & 0.0765 & 0.0876 & 0.2849 & 0.0000 & 0.000 & 0.0000 & 0.000 \\
		75+ & 0.0338 & 0.0024 & 0.0446 & 0.0485 & 0.4106 & 0.0000 & 0.000 & 0.0000 & 0.000 \\
		All ages & 0.0209 & -0.0007 & 0.1209 & 0.1309 & 0.3754 & 0.0000 & 0.000 & 0.0000 & 0.000 \\
		\bottomrule
	\end{tabular}
	
	\flushleft{\footnotesize \sffamily \textit{Note:} Accuracy metrics are aggregated into a mean across all 1,956 areas. }
\end{table}

\begin{table}\caption{Simulation S4: mean accuracy metrics by age group  across $ B=400 $ replications} \label{p3:tab:resultsS4}
	\centering  
	\footnotesize \setlength{\tabcolsep}{5.2pt} 
	\begin{tabular}{lrrrrrrrrr}
		\toprule
		\multirow{2}{0.5cm}{Age (years)}   & \multicolumn{9}{c}{\textbf{S4} \textit{(`HB -- covariates and spatial structure')}}    \\ 
		\cmidrule(lr){2-10} 
		&$ \overline{\text{RMSE}} $& $ \overline{\text{Bias}} $ &$ \overline{\text{ARB}} $ & $ \overline{\text{RSE}}  $& $ \overline{\text{RSEB}}  $& $ \ell(f)$ & $ \overline{\Pr(\ell >0)} $ & $ \hat{\ell}(f) $ & $  \overline{\Pr(\hat{\ell} >0)} $ \\
		\midrule
		\multicolumn{10}{c}{Sampling fraction $ f = 0.02 $ }  \\
		\midrule
		0--14 & 0.0079 & -0.0008 & 0.2072 & 0.2155 & 0.2797 & 0.0000 & 0.000 & 0.0000 & 0.000 \\
		15--29 & 0.0107 & -0.0026 & 0.1845 & 0.1934 & 0.3490 & 0.0000 & 0.000 & 0.0000 & 0.000 \\
		30--44 & 0.0136 & -0.0007 & 0.1187 & 0.1305 & 0.5762 & 0.0000 & 0.000 & 0.0000 & 0.000 \\
		45--59 & 0.0244 & -0.0019 & 0.1039 & 0.1162 & 0.4433 & 0.0000 & 0.007 & 0.0000 & 0.000 \\
		60--74 & 0.0337 & -0.0026 & 0.0750 & 0.0844 & 0.4341 & 0.0000 & 0.000 & 0.0000 & 0.000 \\
		75+ & 0.0345 & 0.0023 & 0.0461 & 0.0495 & 0.4146 & 0.0000 & 0.000 & 0.0000 & 0.000 \\
		All ages & 0.0208 & -0.0010 & 0.1226 & 0.1316 & 0.4162 & 0.0000 & 0.005 & 0.0000 & 0.000 \\
		\midrule
		\multicolumn{10}{c}{Sampling fraction $ f = 0.04 $ }  \\
		\midrule
		0--14 & 0.0077 & -0.0008 & 0.2014 & 0.2097 & 0.2464 & 0.0000 & 0.000 & 0.0000 & 0.000 \\
		15--29 & 0.0104 & -0.0025 & 0.1768 & 0.1866 & 0.2833 & 0.0000 & 0.000 & 0.0000 & 0.000 \\
		30--44 & 0.0132 & -0.0005 & 0.1133 & 0.1265 & 0.4617 & 0.0000 & 0.000 & 0.0000 & 0.000 \\
		45--59 & 0.0234 & -0.0019 & 0.0974 & 0.1113 & 0.3174 & 0.0000 & 0.000 & 0.0000 & 0.000 \\
		60--74 & 0.0321 & -0.0023 & 0.0702 & 0.0807 & 0.3096 & 0.0000 & 0.000 & 0.0000 & 0.000 \\
		75+ & 0.0332 & 0.0021 & 0.0439 & 0.0477 & 0.3318 & 0.0000 & 0.000 & 0.0000 & 0.000 \\
		All ages & 0.0200 & -0.0010 & 0.1172 & 0.1271 & 0.3250 & 0.0000 & 0.000 & 0.0000 & 0.000 \\
		\bottomrule
	\end{tabular}
	
	\flushleft{\footnotesize \sffamily \textit{Note:} Accuracy metrics are aggregated into a mean across all 1,956 areas. }
\end{table}

\newpage
\newpage

\section{Final remarks} \label{p3:sec:conclusion}
This paper has examined the feasibility of a design approach to \ac{SAE} model-based estimation, by proposing a fully Bayesian design algorithm taking a working model and a pilot sample as inputs. The decision rule is based on a new loss function of the sampling fraction: the weighted cell suppression rate. This approach is intended for survey planners and public health professionals to consider an \ac{SAE} problem at the survey design phase, rather than using it strictly for post hoc analyses. The availability of historical survey data is likely to allow survey planners to implement similar approaches at a small cost. The Bayesian decision-analytic approach is well suited to make use of whatever information is available---in this case, a pilot sample---but it is perfectly possible to form a comparable design prior on the basis of good information on model random component (random effect variances, spatial fraction). The \ac{SSD} algorithm then operates on a simulated population that is only as realistic and informative as the design prior is.

Our approach is a direct continuation of previous model-based design decisions considered by \citet{Joseph1995}, \citet{Zou2001} and \citet{Sahu2006}. It establishes a new link to official statistics requirements with a new loss function directly relevant to quality criteria present in most statistical codes of practice internationally. This loss function can be calculated for different social subgroups in situations where estimates of equal precision are needed for minority populations, particularly when equality impact assessments demand equal statistical precision for official statistics across different communities. 

The implementation illustrated in our case study greatly simplifies the design work by expressing the output in terms of effective sampling fraction or \ac{ESS}, which integrates well with existing survey design tools, knowledge around the \ac{DEFF} of various sampling designs, and work to address sample attrition caused by nonresponse.  The case study is designed around random sampling stratified by small area with allocation proportional to size, but more sophisticated approaches can be considered in relation to optimal unequal probability sampling using inferential priority weights as proposed by \citet{Molefe2015}, incorporating on known population characteristics across areas (for instance ethnic diversity or age characteristics). It is also relevant to other model-based estimation techniques such as generalised structure preserving estimation \citep{Zhang2004}.

While the computational burden of this method can remain high in studies involving a large number of cells, the combination of \ac{INLA} and the binary search logic brings these costs to levels that are no longer absurd with modern equipment and high performance cloud computing. Most importantly, the binary search allows planners to anticipate the exact number of iterations needed to reach a prespecified solution precision. They can therefore budget the algorithm's computational requirements.

There are limitations to this approach which make it highly dependent on model misspecification and modelling priors which, together with model selection, arguably constitute the three main challenges to model-based estimation. Given known difficulties in identifying spatial models, it is clear that such a procedure is not likely to be reliable outside of models well established across years of data collection and model testing. While this is not unrealistic with government surveys, this does represent a non-negligible design cost which is difficult to incorporate explicitly in planning.

Our case study conducted on a real population also illustrates that the computation of the loss function being dependent on the reliability of the model, model diagnostics and validation are essential to the reliability and stability of the procedure. Results from the design-based simulation studies show the influence of typical model misspecification on estimation bias and how it clusters in space or social groups. Prior formation also remains, as in many other areas of statistics, a challenge. Recent work in this area by \citet{Fong2010} and \citet{Simpson2014} expresses some of the important obstacles faced by practitioners in this area.

This work tends to support other research showing the feasibility of reconciling Bayesian inference and survey sampling, the intersection of which, in this case, is expressed in terms of \ac{ESS}. This also opens perspectives to combine survey sampling with some very strong developments witnessed in recent years in the design of experiments literature, particularly around adaptive trial designs.

\section*{Acknowledgements}
The author would like to thank Professor Sujit Sahu and Doctor Marta Blangiardo for their helpful comments on an early draft. The author acknowledges the use of the IRIDIS High Performance Computing Facility, and associated support services at the University of Southampton, in the completion of this work.
\section*{Funding}
This work was supported by an Economic \& Social Research Council, Advanced Quantitative Methods doctoral studentship [reference number 1223155].
\section*{Abbreviations}

\begin{acronym} 
	\setlength{\parskip}{0ex}
	\setlength{\itemsep}{1ex}
	\acro{AIC}{Akaike information criterion}
	\acro{AR1}{first-order autoregressive}
	\acro{ARB}{absolute relative bias}
	\acro{AUC}{area under the curve}
	\acro{BMI}{body mass index}
	\acro{CAR}{conditional autoregressive}
	\acro{CCG}{Clinical Commissioning Group}
	\acro{CE}{communal establishment}
	\acro{CSEW}{Crime Survey for England and Wales}
	\acro{CV}{coefficient of variation}
	\acro{DEFF}{design effect}
	\acro{DHS}{Demographic and Health Survey}
	\acro{DIC}{deviance information criterion}
	\acro{EB}{empirical best}
	\acro{EBLUP}{empirical best linear unbiased predictor}
	\acro{EHS}{English Housing Survey}
	\acro{EPP}{empirical plug-in predictor}
	\acro{EPSEM}{equal probability of sampling method}
	\acro{ESS}{effective sample size}
	\acro{FRS}{Family Resources Survey}
	\acro{GIS}{geographical information system}
	\acro{HB}{hierarchical Bayes}
	\acro{HSCIC}{Health \& Social Care Information Centre} 
	\acro{HWB}{Health and Wellbeing Board}
	\acro{ICAR}{intrinsic conditional autoregressive}
	\acro{ICD}{International Classification of Diseases }
	\acro{IHS}{Integrated Household Survey}
	\acro{INLA}{integrated nested Laplace approximation}
	\acro{ISAR}{indirectly standardised emergency hospital admissions rate}
	\acro{JSNA}{joint strategic needs assessment}
	\acro{LAD}{local authority district}
	\acro{LFS}{Labour Force Survey}
	\acro{LISA}{local indicator of autocorrelation}
	\acro{LLB}{Leroux-Lei-Breslow} 
	\acro{LLTI}{limiting long-term illness}
	\acro{LOS}{Life Opportunities Survey}
	\acro{MOR}{median odds ratio}
	\acro{MRP}{multilevel regression and poststratification}
	\acro{MSE}{mean squared error}
	\acro{MSOA}{middle layer super output area}
	\acro{MYPE}{mid-year population estimate}
	\acro{NEX}{normal exchangeable} 
	\acro{NHS}{National Health Service}
	\acro{ONS}{UK Office for National Statistics}
	\acro{PCT}{Primary Care Trust}
	\acro{QLFS}{Quarterly Labour Force Survey}
	\acro{Q--Q}{quantile-quantile}
	\acro{RB}{relative bias}
	\acro{RMSE}{root mean squared error}
	\acro{ROC}{receiver operating characteristics}
	\acro{RSE}{relative standard error}
	\acro{RW1}{first-order random walk}
	\acro{SAE}{small area estimation}
	\acro{SAR}{simultaneous autoregressive}
	\acro{SILC}{Survey on Income and Living Conditions}
	\acro{SMR}{directly age-standardised mortality rate}
	\acro{SRH}{self-rated health}
	\acro{SSD}{sample size determination}
	\acro{SYN}{synthetic predictor}
	\acro{UPSEM}{unequal probability of sampling method}
\end{acronym}

\bibliography{references_NOURL}

\end{document}